%% file: draft.tex
\documentclass[a4paper,11pt]{article}
\pdfoutput=1

\usepackage{jheppub}
\usepackage[T1]{fontenc}
\usepackage{lmodern}

\usepackage{booktabs}
\usepackage{csquotes}
\usepackage{siunitx}
\usepackage{stackengine}
\usepackage[tight]{subfigure}
\usepackage[dvipsnames,table]{xcolor}
\usepackage{xspace}

\input{macros.tex}

\title{NLO QCD and EW corrections to off-shell $\Pt\PZ\Pj$ production at the LHC} 
\author{Ansgar Denner,}
\author{Giovanni Pelliccioli,}
\author{Christopher Schwan}
\affiliation{Institut f\"ur Theoretische Physik und Astrophysik, Universit\"at W\"urzburg, Emil-Hilb-Weg 22, 97074 W\"urzburg, Germany}
\emailAdd{denner@physik.uni-wuerzburg.de}
\emailAdd{giovanni.pelliccioli@physik.uni-wuerzburg.de}
\emailAdd{christopher.schwan@physik.uni-wuerzburg.de}

\abstract{
  The production of a single top quark in association with a $\PZ$
  boson ($\Pt\PZ\Pj$ production)
  at the LHC is a relevant probe of the electroweak
  sector of the Standard Model as well as a window to possible new-physics effects.
  The growing experimental interest in performing differential measurements
  for this process demands an improved theoretical modelling in realistic fiducial
  regions. In this article we present an NLO-accurate $\Pt\PZ\Pj$ calculation that
  includes complete off-shell effects and spin correlations,
  combining QCD and electroweak radiative corrections to the LO signal.
  Integrated and differential cross sections are shown for a fiducial setup
  characterised by three charged leptons, two jets, and missing energy.
}


\begin{document}
\strut\hfill
\maketitle

\section{Introduction}\label{intro}
The production of a single top quark in association with a $\PZ$ boson
is well suited to directly study the electroweak (EW) interactions between two 
of the heaviest particles in the Standard Model (SM).

Differently from processes involving top--anti-top-quark pairs, which are typically
dominated by QCD production mechanisms, single-top processes are mediated by the EW
interaction. Therefore they have smaller cross sections, but also
smaller theoretical uncertainties.
The standard signature for single-top processes is
characterised by a top quark (followed by its leptonic or hadronic decay)
and an additional jet (often labelled as spectator jet), produced in association
with zero, one, or more EW bosons ($\PW,\,\PZ,\,\PH$).

The $\Pt\PZ\Pj$ process is among the very few ones that give direct access to the top-quark
coupling to $\PZ$ bosons. This interaction is poorly known, and a lot of effort
is being put into its investigation both from the experimental and from the theoretical side.
Despite having a similar total cross section as $\Pt\bar{\Pt}\PZ$ production,
$\Pt\PZ\Pj$ production is more suitable to study the $\Pt\bar\Pt\PZ$ coupling as $\Pt\PZ\Pj$ is
an EW process, while $\Pt\bar{\Pt}\PZ$ production is QCD dominated.
Moreover, $\Pt\PZ\Pj$ production gives access to the triple-gauge
($\PW\PW\PZ$) coupling and via the decay of the top quark to the $\Pt\PW\Pb$ coupling.
Owing to the EW-mediated production, the top quark in $\Pt\PZ\Pj$
production is typically polarised,
and therefore the measurements of polarisation-sensitive observables and the extraction of
helicity fractions provide additional probes of the SM and possible
deviations from it \cite{Mahlon:1999gz}.

The importance of $\Pt\PZ\Pj$ as a signal process is shown by several dedicated LHC
measurements performed by CMS and ATLAS with the 13\TeV dataset
\cite{CMS:2017wrv,ATLAS:2017dsm,CMS:2018sgc,ATLAS:2020bhu,CMS:2021ugv}.
The measurement of the total $\Pt\PZ\Pj$ cross section \cite{CMS:2018sgc,ATLAS:2020bhu,CMS:2021ugv},
found to be in good agreement with the SM prediction, represents an
important stress-test of the SM, but performing differential measurements is
expected to give an enhanced sensitivity to possible deviations of top-quark couplings
from their SM values \cite{Degrande:2018fog}. Therefore it is essential that the
theoretical predictions account for the modelling of the decays of the involved resonances.

From the theory side, SM predictions are currently limited to on-shell approximations
for the top-quark description. The next-to-leading-order (NLO) QCD corrections in the SM are
known for many years in the approximation where the production and decay are factorised \cite{Campbell:2013yla}.
Combined NLO EW+QCD corrections in the SM have been computed for an on-shell top quark and off-shell $\PZ$ boson,
including also parton-shower effects \cite{Pagani:2020mov}.
As other single-top processes, $\Pt\PZ\Pj$ production is suited to
compare the five-flavour and four-flavour schemes \cite{Pagani:2020mov} and therefore
to study the b-quark contribution to the proton structure \cite{Campbell:2021qgd}.
Phenomenological investigations of the $\Pt\PZ\Pj$ process have been performed in the
presence of new-physics effects, with
a focus on vector-like top partners \cite{Reuter:2014iya} and anomalous $\Pt\PZ q$ couplings
\cite{Li:2011ek,Kidonakis:2017mfy,Liu:2020bem}. A detailed analysis 
in the SM effective field theory has been carried out in \citeres{Degrande:2018fog,Barman:2022vjd},
where the combination of $\Pt\PZ\Pj$ and $\Pt\PH\Pj$ processes has been shown to enhance the
sensitivity to anomalous values of several SM couplings.

The presented calculation provides the first complete off-shell SM prediction at NLO QCD+EW accuracy
in the five-flavour scheme.
The modelling of the
top-quark and $\PZ$-boson decays accounts for all resonant and non-resonant
contributions and includes complete spin correlations, both at LO and at NLO.
An interesting feature of the off-shell calculation is that,
although at LO the final-state signature selects the decay products of
a (leptonically-decaying) top
quark, the real corrections at NLO (both QCD and EW) inevitably include
partonic processes featuring a (hadronically-decaying) anti-top quark. These contributions, which
are absent in on-shell-approximated calculations, turn out to be
quantitatively important.

The paper is organised as follows. In \refse{details} we describe the
details of our perturbative calculation, the input SM parameters and
the employed fiducial selection cuts, as well as the reconstruction
techniques adopted for the jet and neutrino kinematics. The integrated
cross sections and a number of differential distributions are
discussed in \refse{results}. In \refse{concl} we draw the
conclusions.

\section{Details of the calculation}\label{details}

\subsection{Description of the process}\label{sec:descr}
Following the signal definition of recent LHC analyses \cite{ATLAS:2020bhu,CMS:2021ugv}, we consider the processes
\beq\label{eq:processdef}
\Pp\Pp \rightarrow \Pe^+\Pe^-\mu^+\nu_{\mu}{\rm J}\,{\rm j_{\Pb}}\, + X\,,
\eeq
at NLO EW and QCD accuracy, where ${\rm j_{\Pb}}$ stands for a b jet,
and ${\rm J}$ could be either a light jet or another b jet ($\rm J=j_b,j$).
In the five-flavour scheme, the LO process receives contributions from partonic channels that only
involve quarks as external coloured particles:
\begin{equation*}
  \bar{q}_{\Pd}\, q_{\Pu}\rightarrow\bar{\Pb}\,\Pb\ \Pe^+ \Pe^- \mu^+ \nu_\mu
  \,,\qquad
  q_{\Pu}\overset{\scriptscriptstyle(-)}{\Pb} 
  \rightarrow {q}_{\Pd} \overset{\scriptscriptstyle(-)}{\Pb} \ \Pe^+ \Pe^- \mu^+ \nu_\mu
  \,,\qquad
  \bar{q}_{\Pd}\overset{\scriptscriptstyle(-)}{\Pb} \rightarrow \bar{q}_{\Pu}
  \overset{\scriptscriptstyle(-)}{\Pb} \ \Pe^+ \Pe^- \mu^+ \nu_\mu\,.
\end{equation*}
At cross-section level, three tree-level perturbative orders are present,
namely  $\mc O(\alpha^6)$, $\mc O(\as^2\alpha^4)$, and the
interference $\mc O(\as\alpha^5)$. However, the EW production of a top quark
and a $\PZ$ boson can only take place at $\mc O(\alpha^6)$, which is in fact
regarded as the LO signal.
The interference, of $\mc O(\as\alpha^5)$, vanishes due to colour algebra,
since the mixing of the bottom quark with
the light quarks is neglected (a unit CKM matrix is assumed).
The $\mc O(\as^2\alpha^4)$ contributions and NLO corrections on top of them are not
considered in this paper.

\begin{figure}[tb]
  \centering
  \subfigure[$t$ channel\label{fig:lo_t}]{\includegraphics[scale=0.45]{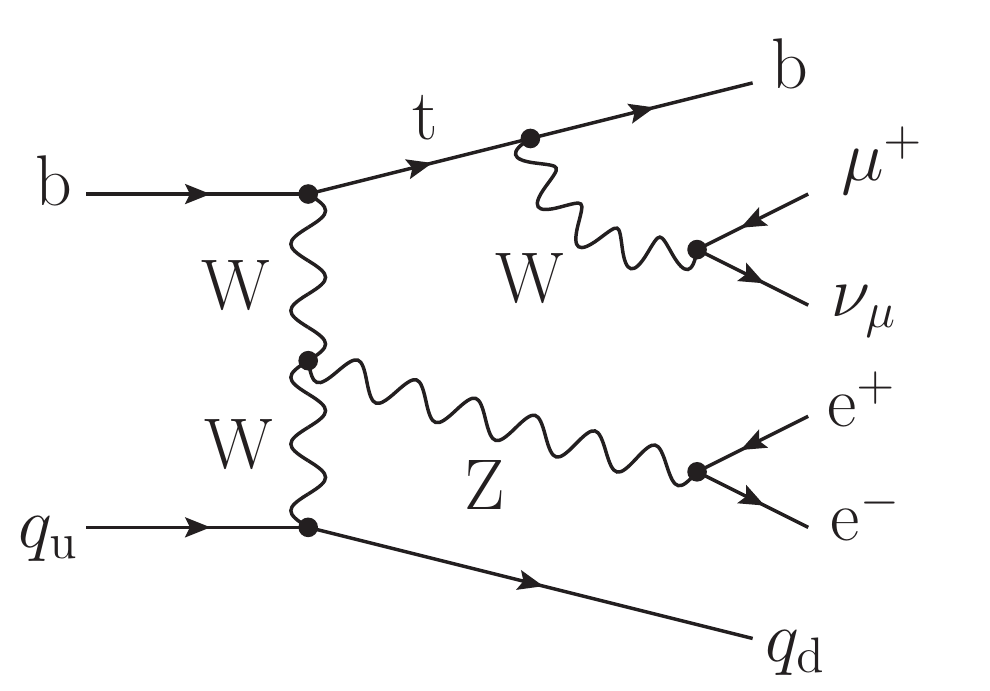}}
  \subfigure[$s$ channel\label{fig:lo_s}]{\includegraphics[scale=0.45]{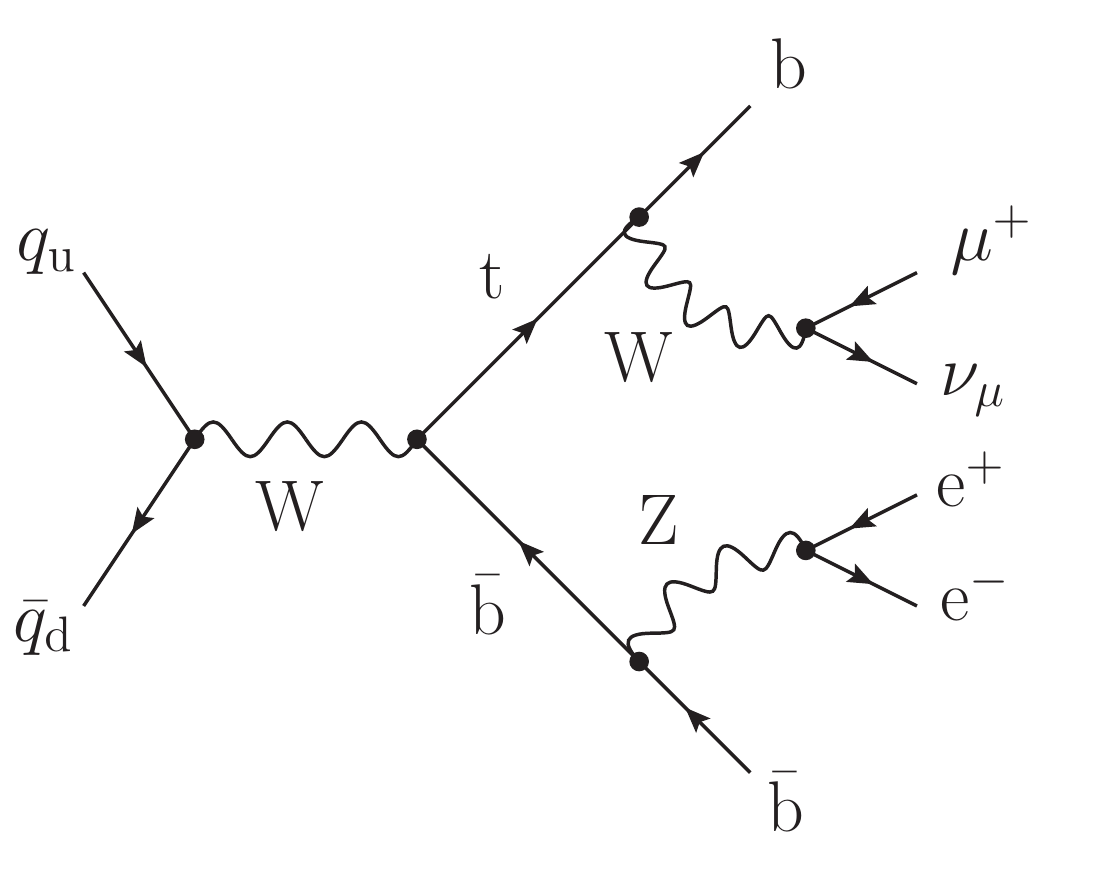}}\\
  \subfigure[Non resonant\label{fig:lo_nr}]{\includegraphics[scale=0.45]{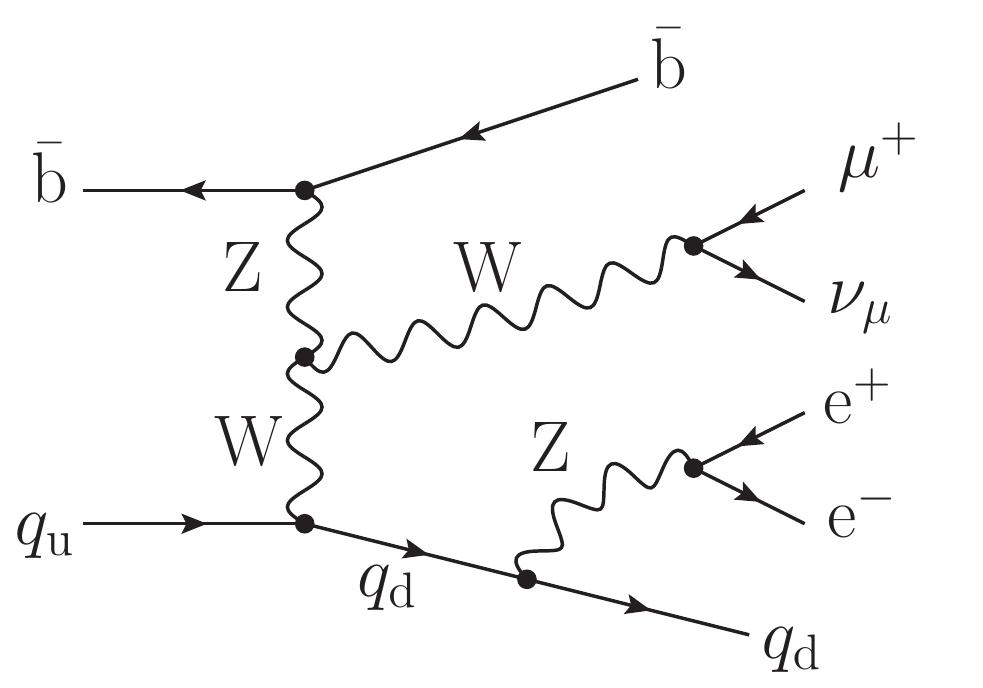}}
  \subfigure[Vector-boson scattering\label{fig:lo_vbs}]{\includegraphics[scale=0.45]{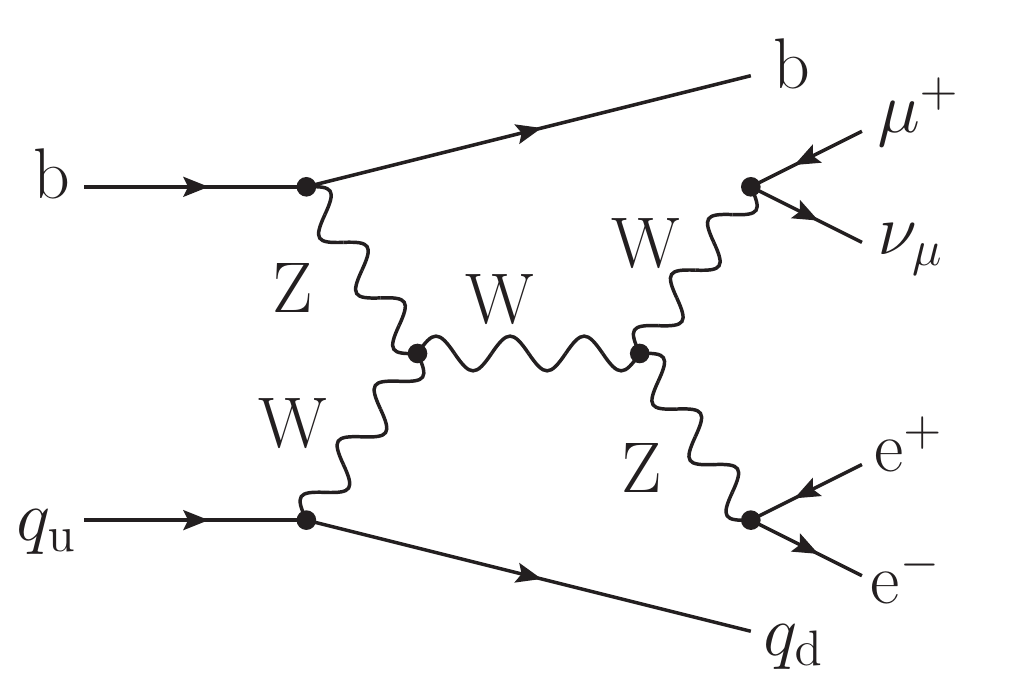}}
  \caption{Sample tree-level diagrams contributing at
    $\mc O(\alpha^6)$ to off-shell $\Pt\PZ\Pj$ production at the LHC.}
  \label{fig:lodiags}
\end{figure}

With the signal definition of Eq.~\eqref{eq:processdef}, it is easy to see that
the production of an (off-shell) top quark can take place both in $s$ channel
($\bar{q}_{\Pd}\,q_{\Pu}$ initial state) and in $t$ channel ($q_{\Pu}\overset{\scriptscriptstyle(-)}{\Pb},\, \bar{q}_{\Pd}\overset{\scriptscriptstyle(-)}{\Pb}$ initial states).
Sample diagrams are shown in \reffi{fig:lodiags}.
It is essential to recall that at LO a clear distinction between $s$- and $t$-channel
contributions is possible, owing to a different number of b quarks in the final state
[see \reffis{fig:lo_t}--\ref{fig:lo_s}]. However, starting from NLO corrections (both QCD and EW),
such a separation between the two top-quark production mechanisms is ill defined, \ie
the different contributions are not separately gauge invariant
owing to partonic channels that embed both $s$- and $t$-channel contributions, as in
off-shell single-top production \cite{Falgari:2011qa,Frederix:2019ubd}.
All resonant and non-resonant  [\reffi{fig:lo_nr}] contributions are included for all partonic channels.
Contributions without a top-quark resonance as those embedding the
vector-boson scattering subprocess [\reffi{fig:lo_vbs}] are expected to be
sub-dominant w.r.t. the top-quark-resonant ones.

At NLO there are four different perturbative orders, but in this paper we only
consider corrections of $\mc O(\alpha^7)$ and $\mc O(\as\alpha^6)$.
The former are genuine EW corrections to the leading EW order.
The latter naively include two kinds of corrections: the QCD corrections
to the leading EW order and the EW ones to the LO interference.
However, since virtual or real EW corrections
do not change the vanishing colour structure of the LO interference, the $\mc O(\as\alpha^6)$
is only made of pure QCD corrections to the LO EW contribution.

The following real partonic processes contribute at $\mc O(\as\alpha^6)$:
\begin{align*}
  \bar{q}_{\Pd}\,q_{\Pu}& \rightarrow\bar{\Pb}\,\Pb\ \Pg\ \Pe^+ \Pe^- \mu^+ \nu_\mu\,,\quad
  q_{\Pu}\overset{\scriptscriptstyle(-)}{\Pb} \rightarrow {q}_{\Pd} \overset{\scriptscriptstyle(-)}{\Pb}
  \ \Pg\ \Pe^+ \Pe^- \mu^+ \nu_\mu\,,\quad
  \bar{q}_{\Pd}\overset{\scriptscriptstyle(-)}{\Pb} \rightarrow \bar{q}_{\Pu} \overset{\scriptscriptstyle(-)}{\Pb} \ \Pg\ \Pe^+ \Pe^- \mu^+ \nu_\mu\,,\\
  \Pg\,q_{\Pu}& \rightarrow\bar{\Pb}\,\Pb\ {q}_{\Pd}\ \Pe^+ \Pe^- \mu^+ \nu_\mu\,,\quad
  \bar{q}_{\Pd}\,\Pg\rightarrow\bar{\Pb}\,\Pb\ \bar{q}_{\Pu}\ \Pe^+ \Pe^- \mu^+ \nu_\mu\,,\quad
  \,\,\Pg\ \overset{\scriptscriptstyle(-)}{\Pb} \rightarrow \bar{q}_{\Pu} \,{q}_{\Pd} \,\overset{\scriptscriptstyle(-)}{\Pb} \ \Pe^+ \Pe^- \mu^+ \nu_\mu\,.
\end{align*}
The same processes contribute at $\mc O(\alpha^7)$, upon replacing external gluons with photons.
The gluon-induced channels that open up at NLO QCD give a sizeable contribution owing to the enhancement
from the large gluon luminosity in the proton. In contrast, the photon-induced real corrections are
suppressed by coupling power counting $[\mc{O}(\alpha/\as)]$ and by the small photon luminosity in the proton.

\begin{figure}[tb]
  \centering
  \includegraphics[scale=0.49]{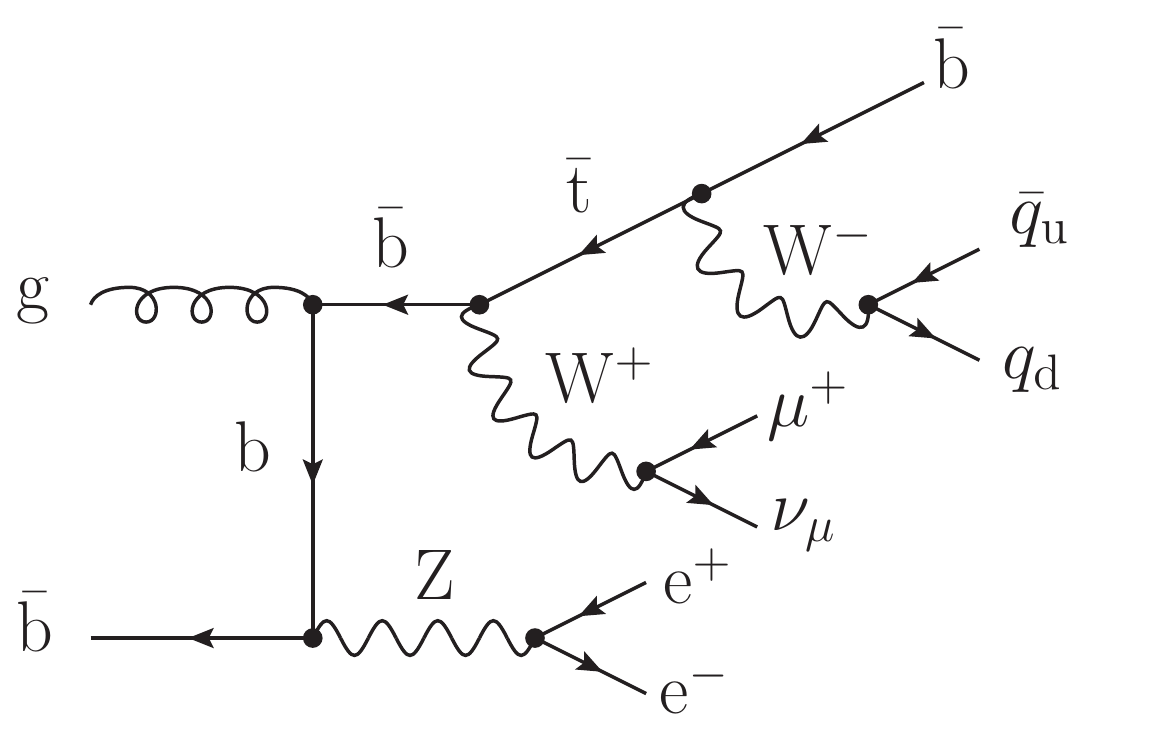}
  \caption{Sample diagram for anti-top-quark production in the $\Pg \bar{\Pb}$ channel at $\mc O(\as\alpha^6)$.}
  \label{fig:antitopdiag}
\end{figure}

The new partonic channels that contribute at NLO can also enhance the cross section
due to different underlying resonance structures with respect to those present at LO. In particular,
the processes
\begin{equation}
\bar{\Pb}\, \Pg,\, \bar{\Pb}\, \gamma  \rightarrow \bar{q}_{\Pu} \,{q}_{\Pd} \,\bar{\Pb} \ \Pe^+ \Pe^- \mu^+ \nu_\mu
\label{eq:anti-top-production}
\end{equation}
allow for the production of a resonant anti-top quark followed by its hadronic decay
($\bar{\Pt}\rightarrow \bar{q}_{\Pu} \,{q}_{\Pd} \,\bar{\Pb}$), as shown in the sample diagram in
\reffi{fig:antitopdiag}.
Such a contribution, which is absent in on-shell-approximated calculations \cite{Campbell:2013yla,Pagani:2020mov},
is non-negligible and could be suppressed using a jet veto requiring at most one light jet.
Since the same considerations hold for the charge-conjugated process of Eq.~\eqref{eq:processdef},
if both ${\Pt}\PZ\Pj$ and $\bar{\Pt}\PZ\Pj$ production
were included in the signature as in experimental analyses
\cite{CMS:2017wrv,ATLAS:2017dsm,CMS:2018sgc,ATLAS:2020bhu,CMS:2021ugv},
\beq
\Pp\Pp \rightarrow \Pe^+\Pe^-
\mu^\pm  \overset{\scriptscriptstyle(-)}{\nu}_{\!\!\mu\,}
{\rm J}\,{\rm j_{\Pb}}\, + X\,,
\eeq
the contributions from $\bar{\Pt}\PW^+\PZ$ and ${\Pt}\PW^-\PZ$
intermediate states would both
give a similar relative correction to the respective cross section.

\begin{figure}[tb]
  \centering
  \subfigure[$\Pg\rightarrow\Pb\bar{\Pb}$\label{fig:udx}]{\includegraphics[scale=0.45]{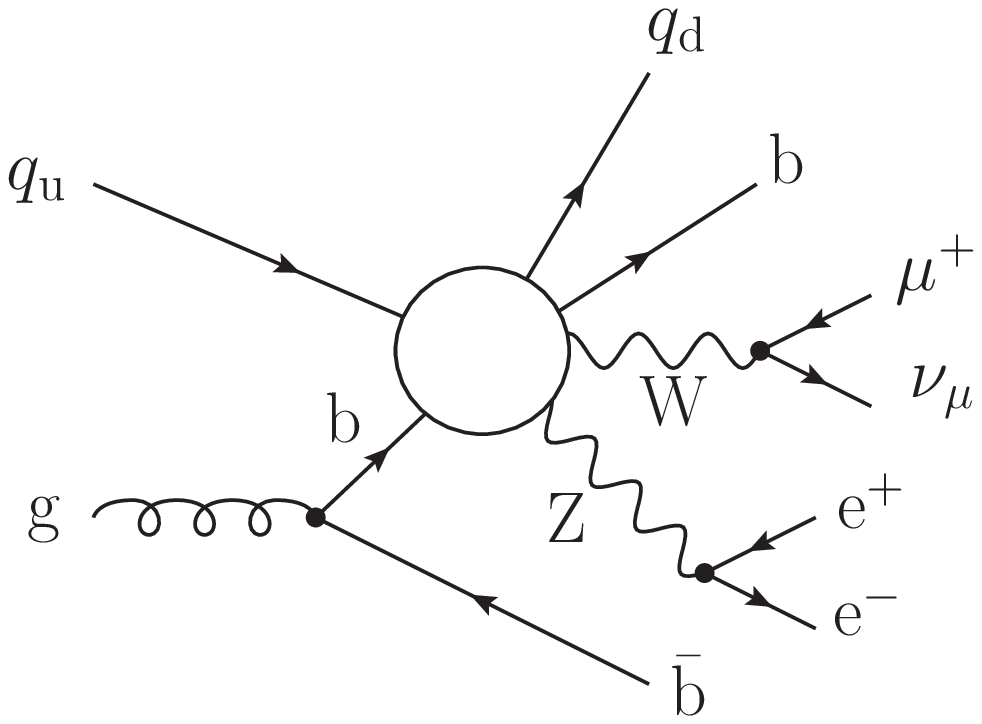}}
  \subfigure[$\Pg\rightarrow\bar{\Pb}\Pb$\label{fig:ub}]{\includegraphics[scale=0.45]{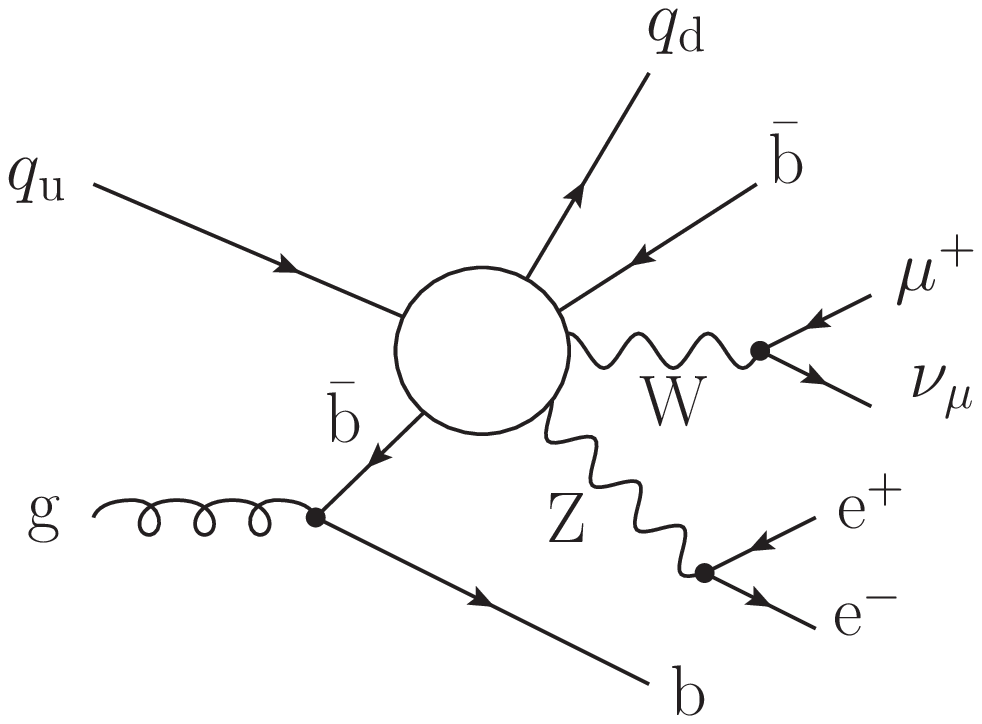}}
  \subfigure[$\Pg\rightarrow\bar{q}_{\Pd}q_{\Pd}$\label{fig:ubx}]{\includegraphics[scale=0.45]{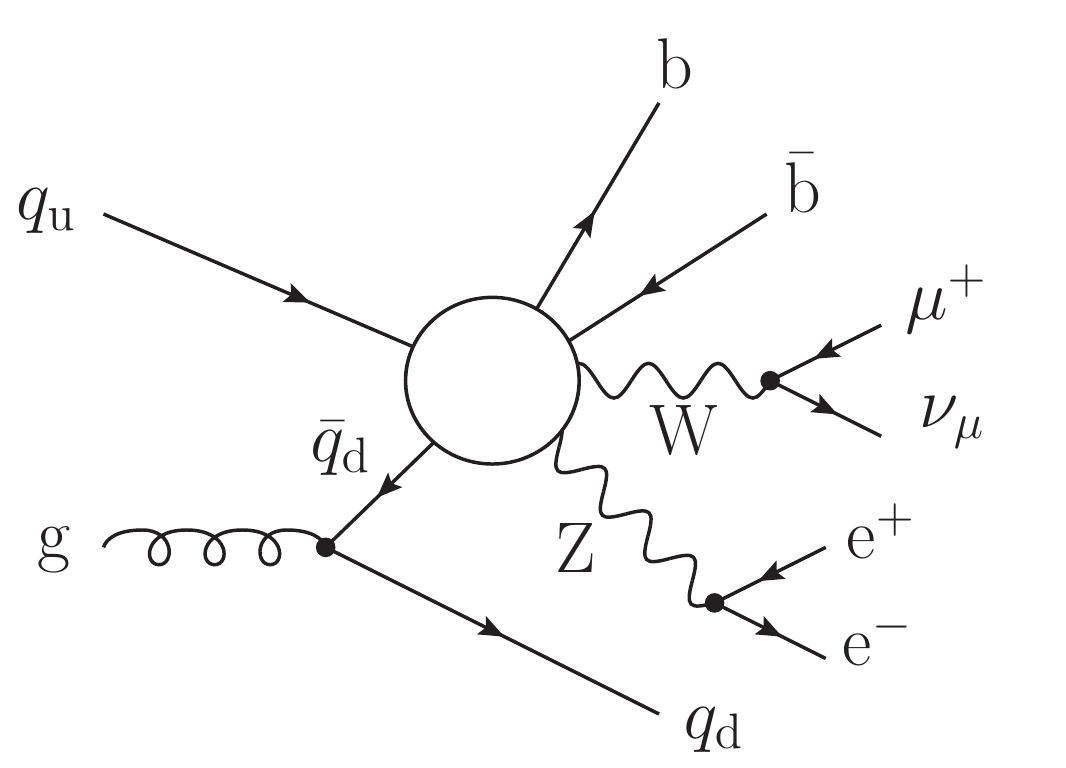}}
  \caption{Real-radiation topologies embedding different initial-state
    collinear splittings in the $\Pg q_{\Pu}$ channel at
    $\mc O(\as\alpha^6)$.}
  \label{fig:nloRdiags}
\end{figure}

The analysis of the initial-state singularities in gluon-induced channels also shows that a
distinction between $s$- and $t$-channel contributions is not well defined at NLO. 
In \reffi{fig:nloRdiags} we highlight the underlying Born-level processes
that are embedded in the real-radiation channel $\Pg q_{\Pu}$.
Comparing the three topologies it is clear that the same real partonic process may
include both $s$- and $t$-channel single-top diagrams, which are not separately
gauge-invariant.

A large number of soft- and collinear-singular configurations need to be
subtracted in order to render the calculation infrared safe, both at NLO QCD
and at NLO EW. The number of subtraction terms is especially
large in the $\mc{O}(\alpha^7)$ corrections due to the presence of seven external
charged particles at LO, while the number of subtraction terms needed at $\mc{O}(\as\alpha^6)$
is smaller owing to only four coloured external particles at Born level.
It is worth noticing that, thanks to the requirement of at least one
b~jet in the final state, it is not
necessary to include subtraction terms for a $\gamma\rightarrow \bar\Pb\Pb$
splitting in the final state, which appears in gluon-induced real contributions. The collinear singularity associated with this splitting
is in fact cut out thanks to a $ \bar{\Pb}{\Pb} \rightarrow {\rm j}$ recombination
condition that we apply to make the jet algorithm infrared safe also in the
presence of flavoured jets \cite{Banfi:2006hf}. The same argument holds at $\mc O(\alpha^7)$
for photon--quark-induced partonic channels.
See \refse{sec:select}, in particular Eq.~\refeq{eq:reco_rules}, for a more detailed discussion of the recombination algorithm.

The one-loop amplitudes that enter the virtual corrections both at NLO EW and at NLO QCD
involve up to 8-point functions. However, in spite of the presence
of several mass scales making the loop-integral evaluation more involved, the
virtual corrections ($2\rightarrow 6$ process) are computationally less expensive than the real
corrections ($2\rightarrow 7$ process).

The calculation is performed within the MoCaNLO Monte Carlo framework.
MoCaNLO has already been used to compute full NLO corrections to LHC processes
with a high number of particles in the final state, including underlying
resonance structures with top quarks
\cite{Denner:2015yca,Denner:2016jyo,Denner:2016wet,Denner:2017kzu,Denner:2020hgg,Denner:2021hqi}.
MoCaNLO relies on tree-level and one-loop SM amplitudes provided by \recola
\cite{Actis:2012qn, Actis:2016mpe} 
and computed with the help of the \collier library  \cite{Denner:2016kdg}
 for tensor-integral reduction \cite{Denner:2002ii,Denner:2005nn} and loop-integral evaluation \cite{Denner:2010tr}.
The dipole formalism \cite{Catani:1996vz,Dittmaier:1999mb,Catani:2002hc,Dittmaier:2008md} is employed to take care of the cancellation of
soft and collinear singularities of QCD and QED origin. The $\MSbar$ factorisation scheme
is used for the treatment of initial-state collinear singularities.

The presented calculation of the NLO QCD + EW corrections to
$\Pt\PZ\Pj$ production can be viewed as
complementary to the one
of $\PW\PZ$ scattering \cite{Denner:2019tmn}, which has also been carried out in the MoCaNLO framework.
The two calculations basically share the same final state [see \reffi{fig:lo_vbs}], the only difference being the
minimum number of required b-tagged jets.
However, the dominant resonance structure is different in the two processes,
which motivates
rather different kinematic selections and well-distinguished signal definitions in LHC analyses.

In this context we only focus on the perturbative orders that enable the presence
of a top quark as underlying resonance, \ie the EW signal, while we ignore the
QCD background and NLO corrections to it. In spite of the absence of a resonant top quark,
the QCD background ($\PW\PZ+$jets) is expected to be sizeable and thus deserves
a tailored phenomenological study.

As a last comment of this section, we stress that all results presented in this paper
are within the five-flavour scheme. This scheme has the advantage of resumming effectively
the logarithms of the type $\as^n \log^m({\mu^2/m_\Pb^2})$ for $m\leq n$, but the disadvantage
of having at NLO QCD a leading-order-like renormalisation-scale dependence.
Since the typical hard scale $\mu$ of the process is much larger than the $\Pb$-quark mass,
the power corrections of type $(m_{\Pb}/\mu)^n$ can be safely neglected.
A four-flavour calculation would include the complete $m_{\Pb}$ dependence and provide an 
actual NLO scale dependence
on the renormalisation scale when including QCD corrections \cite{Pagani:2020mov}.
In the four-flavour scheme
gluon-initiated processes contribute already at LO, resulting in sizeable
$\Pt\bar{\Pt}\PZ$ and $\Pt\PW\PZ$ contamination, which needs to be taken care by specific
cuts and vetoes. Although a comparison between the two schemes including off-shell effects (see
\citere{Pagani:2020mov} for on-shell top quarks) is desirable,
this is beyond the scope of this work. 

\subsection{Input parameters}\label{sec:input}
We compute LO and NLO cross sections in the SM for the process defined in Eq.~\eqref{eq:processdef}.
All leptons are considered massless. The five-flavour scheme ($N_{\text{f}}=5$) is employed,
and we therefore include contributions with initial-state b~quarks,
which are assumed to be massless as the other light quarks.
A unit CKM matrix is used, leading to no mixing between different quark families.
The values for the on-shell masses and widths of EW bosons
are \cite{ParticleDataGroup:2020ssz},
\begin{equation}
\begin{aligned}
\Mwo &= 80.379 \GeV,\qquad &  \Gwo &= 2.085\GeV, \\
\Mzo &= 91.1876 \GeV,\qquad & \Gzo &= 2.4952\GeV.
\end{aligned}
\end{equation}
The on-shell values are converted into the corresponding pole values
according to the relations \cite{Bardin:1988xt}
\beq
M_V =\frac{M_V^{\rm OS}}{\sqrt{1+\left(\Gamma_V^{\rm OS}/M_V^{\rm OS}\right)^2}}\,,\qquad
\Gamma_V =\frac{\Gamma_V^{\rm OS}}{\sqrt{1+\left(\Gamma_V^{\rm OS}/M_V^{\rm OS}\right)^2}}\,.
\eeq
The mass and width of the Higgs boson are set to the following values \cite{ParticleDataGroup:2020ssz},
\begin{align}
\MH  ={}& 125 \GeV, \qquad \GH= 4.07\,\MeV.
\end{align}
The NLO width of the top quark is used in all contributions to the cross section.
We apply relative NLO EW and QCD corrections from \citere{Basso:2015gca} to the LO top-quark width
computed following \citere{Jezabek:1988iv}.
The numerical values read
\begin{align}
\Mt = 173\GeV,\qquad
\Gt
= 1.3636\GeV.
\end{align}
The EW coupling is defined in the $G_\mu$~scheme \cite{Denner:2000bj}, with the Fermi constant set to
\beq
G_\mu = 1.16638\cdot10^{-5} \GeV^{-2}.
\eeq
EW-boson and top-quark masses, as well as of the EW mixing angle, are treated within the complex-mass scheme
\cite{Denner:1999gp,Denner:2000bj,Denner:2005fg,Denner:2006ic,Denner:2019vbn}.

The evaluation of parton-distribution functions (PDFs) and the running of $\as$ are performed with
the {\scshape LHAPDF6} interface \cite{Buckley:2014ana}.
We use the \sloppy \texttt{NNPDF31\_nnlo\_as\_0118\_luxqed} PDF set
\cite{Bertone:2017bme} throughout the calculation.

The renormalisation and factorisation scales are simultaneously set to the following dynamical value,
\beq
\mu_{\rm R}^0 = \mu_{\rm F}^0 =\frac{M_{\rT, \Pt}+M_{\rT, \PZ}}6\,,
\label{eq:central-scale-choice}
\eeq
adapting the choice of \citere{Pagani:2020mov} to the inclusion of decay products of the
top quark.
In particular, we define the transverse mass of the $\PZ$ boson
as
\beq
M_{\rT, \PZ}=\sqrt{\MZ^2+
  \left(\vec{p}_{\rm T, e^+}+\vec{p}_{\rm T, e^-}\right)^2}\,.
\eeq
If both tagged jets are b~jets, the top-quark momentum is reconstructed
with the b~jet ($\rm j_b^{\rm best}$) that gives an invariant mass of the $\mu^+\nu_\mu \rm j_b$ system
closest to the top-quark pole mass. If only one tagged b~jet is present, it is automatically associated
to the top quark, in a formula,
\beq\label{eq:mttop}
M_{\rT, \Pt}=\sqrt{\Mt^2+
   \left(\vec{p}_{\rm T, \mu^+}+\vec{p}_{\rm T, \nu_\mu}+\vec{p}_{\rm T, j_{b}^{\rm best}}\right)^2}\,.
\eeq
When performing scale variations we vary $\mu_\mathrm{R} = \xi_\mathrm{R} \mu_\mathrm{R}^0$ and $\mu_\mathrm{F} = \xi_\mathrm{F} \mu_\mathrm{F}^0$ around the central choice $\mu_\mathrm{R}^0$ and $\mu_\mathrm{F}^0$ by the following factors $(\xi_\mathrm{R}, \xi_\mathrm{F})$:
\begin{equation}
\bigl( 1/2, 1/2 \bigr) \text{,} \quad
\bigl(   1,   1 \bigr) \text{,} \quad
\bigl(   2,   2 \bigr) \text{,} \quad
\bigl(   1, 1/2 \bigr) \text{,} \quad
\bigl( 1/2, 1   \bigr) \text{,} \quad
\bigl(   1,   2 \bigr) \text{,} \quad
\bigl(   2,   1 \bigr) \text{,}
\label{eq:scale-variations}
\end{equation}
and take the envelope (minimum and maximum value of the varied cross sections) to estimate the QCD uncertainty.

\subsection{Selection cuts}\label{sec:select}
Only particles (charged under QED or QCD) with $|y|<5$ undergo jet clustering, which is achieved
using the $k_{\rT}$ jet algorithm \cite{Catani:1991hj,Catani:1993hr,Ellis:1993tq} with resolution
radius $R=0.4$ and $R=0.1$, respectively, for jets (both at NLO QCD and NLO EW) and dressed leptons (at NLO EW).
The $k_{\rT}$ jet algorithm is chosen over the corresponding anti-$k_{\rT}$ algorithm since the former
is known\footnote{We note that during the writing of this paper two articles have appeared that propose
the extension of the anti-$k_{\rT}$~\cite{Czakon:2022wam} and even general~\cite{Gauld:2022lem}
jet algorithms to safely define flavoured jets.} to have an infrared-safe definition \cite{Banfi:2006hf}
at all orders in the presence of flavoured jets.
Thus, we label jets as either light jets ($\Pj$), or $\Pb$~jets ($\mathrm{j}_{\Pb}$) using the
following recombination rules:
\begin{equation}\label{eq:reco_rules}
\overset{\scriptscriptstyle(-)}{\Pb}\Pg     \rightarrow {\Pj}_{\Pb} \text{,}  \quad
\overset{\scriptscriptstyle(-)}{\Pb}\gamma  \rightarrow {\Pj}_{\Pb} \text{,}  \quad
\bar\Pb\,\Pb                                \rightarrow {\Pj}       \text{,} \quad
\Pg\,\Pg                                    \rightarrow {\Pj}       \text{,}  \quad
\overset{\scriptscriptstyle(-)}{q}\Pg       \rightarrow {\Pj}       \text{,}  \quad
\overset{\scriptscriptstyle(-)}{q}\gamma    \rightarrow {\Pj}       \text{,} \quad
\ell^\pm\gamma                             \rightarrow {\ell^\pm}    \text{.}
\end{equation}
As introduced in \refse{sec:descr} we follow the setup of the recent ATLAS analysis \cite{ATLAS:2020bhu}.
We ask for events with
\beq\label{cuts:njets}
N_{\rm J}=N_{\rm j}+N_{\rm j_{\Pb}}\geq 2\,,\qquad N_{\rm j_{\Pb}}\geq 1\,,
\eeq
and assume 100\% b-tagging efficiency. Jets are required to satisfy
\beq \label{cuts:ptjets}
\pt{\rm j}> 35\GeV,\qquad\pt{\rm j_b} > 35\GeV, \qquad |y_{\rm j} |<4.5\,,
\qquad |y_{\rm j_{\Pb}} |<2.5\,.
\eeq
Furthermore, we ask for exactly three charged leptons ($\Pe^+\Pe^-\mu^+$) with
\beq
  |y_{\ell^\pm} |<2.5\,,\qquad \pt{\ell_1}> 28\GeV,\qquad
  \pt{\ell_{2}},\,\pt{\ell_{3}} > 20\GeV, 
\eeq
where the charged leptons are sorted according to their transverse momentum,
$\ell_1$ being the leading one.
The invariant mass of the opposite-sign, same-flavour lepton pair
($\Pe^+\Pe^-$ in our case) is constrained by setting
\beq
M_{\Pe^+\Pe^-}>30\GeV.
\eeq
A minimum distance is required between jets and charged leptons,
\beq\label{cuts:jl}
\Delta R_{\rm j,\ell}>0.4\,,\qquad \Delta R_{\rm j_{\Pb},\ell}>0.4\,.
\eeq

Besides the default setup just described, we consider a setup
with an additional cut on the invariant mass of the $\Pe^+\Pe^-$ pair
\begin{equation}
\SI{81}{\giga\electronvolt} < M_{\mathrm{e}^+\mathrm{e}^-} < \SI{101}{\giga\electronvolt} \text{.}
\label{eq:invariant-mass-cut}
\end{equation}
This cut selects events close to the $\PZ$-boson resonance and is therefore
called \emph{Z-peak setup}.


\subsection{Event topologies and kinematic reconstruction}
\label{sec:kinreco}

Since we consider final states with more than one b-flavoured jet, the identification of the
b~jet from the top-quark decay is ambiguous. To properly define physical observables, we need
to solve this ambiguity by means of some discrimination criterion.

In the considered setups, at least two jets are required to fulfil the selections described in
Eqs.~\eqref{cuts:ptjets} and \eqref{cuts:jl}. At LO only two-jet events
are present. At NLO, the additional QCD or photon radiation may be clustered with other partons (giving a
two-jet event) or result in a third jet. If also the third jet fulfils the requirements of
Eqs.~\eqref{cuts:ptjets} and \eqref{cuts:jl}, we have a three-jet event,
otherwise we have a two-jet event.

The definitions we use for the \emph{top-decay jet} ($\tj$) and for the \emph{spectator jet} ($\sj$)
are inspired by single-top NLO studies \cite{Cao:2005pq,Schwienhorst:2010je} and mimic
those used in the most recent CMS analysis~\cite{CMS:2021ugv}.

In the case of a two-jet event with one b~jet and one light jet, there is no ambiguity, therefore
the b~jet is labelled as the top-decay jet, while the light jet is labelled as the  spectator jet.
If both jets are b~jets, the {top-decay jet} is the one that gives an invariant mass closest
to the top-quark pole mass when combined with the top-decay lepton ($\mu^+$  in our setup)
and with the reconstructed neutrino (see below). The other b~jet is labelled as the {spectator jet}.

In the case of a three-jet event with one b~jet and two light jets,
the b~jet is labelled as the {top-decay jet} and the
hardest-$p_{\rm T}$ light jet is labelled as the {spectator jet}.
If there are two b~jets and one light jet, the {top-decay jet} is the b~jet that gives the
closest invariant mass to the top-quark pole mass when combined with the top-decay lepton
and the reconstructed neutrino. The {spectator jet} is chosen as the light jet.

The reconstruction of the neutrino is performed with the top-resonance-aware method
used in \citere{CMS:2021ugv}.
The missing transverse momentum, $\pt{\rm miss}$, is assumed to be the transverse momentum of the only neutrino present in the event (always the case for the signal process). The longitudinal component of the neutrino momentum is reconstructed imposing $(p_{\mu^+}+p^{\rm rec}_{\nu})^2=\MW^2$:
if the solutions of the quadratic equation are complex, the real part
is selected; if two real solutions exist, the solution which minimises
$|(p_{\tj}+p_{\mu^+}+p^{\rm rec}_{\nu})^2-\Mt^2|$, where
$p_{\tj},\,p_{\mu^+}$ and $p^{\rm rec}_{\nu}$ are the momenta of the
top-decay jet, the anti-muon and reconstructed neutrino, respectively,
is taken.

If a two-fold ambiguity is present both in defining the top-decay jet, \ie in
events with two b~jets, and in neutrino reconstruction, \ie two real solutions
of the quadratic equation for the on-shell requirement, the minimisation of
$|(p_{\tj}+p_{\mu^+}+p^{\rm rec}_{\nu})^2-\Mt^2|$ is performed over the four possible
combinations of $p_{\tj}$ and $p^{\rm rec}_{\nu}$.

\subsection{Validation}\label{valid}
The correct implementation of infrared subtraction terms has been
thoroughly checked by varying the dipole parameters $\alpha_{\rm dip}$ \cite{Catani:2002hc}
which control the correct subtraction of infrared-singular configurations between subtraction
dipoles and the corresponding integrated counterparts. The complete calculation of $\Pt\PZ\Pj$
production has been performed both with $\alpha_{\rm dip}=10^{-2}$ and with $\alpha_{\rm dip}=1$,
finding good agreement within the uncertainties of the Monte Carlo integration.

In order to further check the cancellation of infrared poles
in the sum of virtual contributions and of $I$ operators in integrated dipoles
\cite{Catani:1996vz,Catani:2002hc}, we have also performed variations of the infrared scale
$\mu_{\rm IR}$ that appears in the logarithms multiplying single and double poles of virtual
origin (both in QED and in QCD) in dimensional
regularisation.
Agreement within integration errors has been found at integrated level using
$\mu_{\rm IR}=\Mt/10,\,\Mt,\,10^4\,\Mt$.

\section{Results}
\label{results}

We present integrated cross sections in \refse{sec:integrated-cross-sections}, compare them to results from the literature in \refse{sec:comparison-with-literature-results}, and discuss differential distributions in \refse{sec:differential-distributions}.

\subsection{Integrated cross sections}
\label{sec:integrated-cross-sections}

\begin{table}
\centering
\def\stackalignment{r}
\newcommand{\asyunc}[2]{{\scriptsize\stackanchor[1pt]{\SI[retain-explicit-plus=true]{#2}{\percent}}{\SI{#1}{\percent}}}}
\begin{tabular}{lS[table-format=3.3(2)]@{}rS[table-format=3.1]S[table-format=3.3(2)]@{}rS[table-format=3.1]}
\toprule
             & \multicolumn{3}{c}{Default setup}
             & \multicolumn{3}{c}{Z-peak setup} \\
Contribution & \multicolumn{2}{c}{$\sigma$}           & $\delta$
             & \multicolumn{2}{c}{$\sigma$}           & $\delta$ \\
             & \multicolumn{2}{c}{[\si{\femto\barn}]} & [\si{\percent}]
             & \multicolumn{2}{c}{[\si{\femto\barn}]} & [\si{\percent}] \\
\midrule
$\mathcal{O} (\alpha^6) = \text{LO}$ &
    0.6415(0) & \asyunc{-13.5}{+8.9} & 100.0 & 0.5846(0) & \asyunc{-13.5}{+9.0} & 100.0 \\
$\mathcal{O} (\as \alpha^6)$ &
    0.1988(4) & & 31.0 & 0.1788(3) & & 30.6 \\
$\mathcal{O} (\alpha^7)$ &
    -0.0414(2) & & -6.4 & -0.0497(3) & & -8.5 \\
\midrule
NLO QCD    & 0.8403(4) & \asyunc{-3.9}{+8.6} & 131.0 & 0.7634(3) & \asyunc{-3.9}{+8.6} & 130.6 \\
NLO EW     & 0.6002(2) & \asyunc{-13.9}{+9.4} & 93.6 & 0.5349(3) & \asyunc{-13.9}{+9.4} & 91.5 \\
NLO QCD+EW & 0.7990(4) & \asyunc{-4.2}{+9.4} & 124.5 & 0.7137(4) & \asyunc{-4.4}{+9.8} & 122.1 \\
\bottomrule
\end{tabular}
\caption{Integrated cross sections, $\sigma$, for LO, NLO QCD, NLO EW, and NLO QCD+EW
  in the default setup and the Z-peak setup [with the additional cut Eq.~\eqref{eq:invariant-mass-cut}].
Numbers in parentheses are statistical uncertainties from the Monte Carlo integration.
Zero entries indicate uncertainties smaller than the given precision.
Asymmetric percentages give the envelope of a 7-point scale variation.
Finally, $\delta$ is the size of each contribution in terms of the LO.}
\label{tab:integrated-cross-sections}
\end{table}
In \refta{tab:integrated-cross-sections} we show the integrated
cross sections for LO, NLO QCD, NLO EW, and NLO QCD+EW, together with
the corresponding QCD scale variation and the size of
corrections/contributions relative to the LO cross section
$\sigma^\text{LO}$ of $\mathcal{O} (\alpha^6)$,
\begin{equation}
\delta = \frac{\sigma}{\sigma^\text{LO}} \text{,}
\label{eq:delta-definition}
\end{equation}
for the two different setups described in \refse{sec:select}, the
default setup and the Z-peak setup.
\begin{table}%
\centering
\setlength{\tabcolsep}{5pt} 
\begin{tabular}{rS[table-format=3.1]S[table-format=2.2]S[table-format=2.1]S[table-format=2.2]S[table-format=3.1]S[table-format=4.1]S[table-format=4.1]S[table-format=4.1]}
\toprule
ch. & $\delta_\text{sum}^\text{LO}$
    & $\delta_\text{sum}^\text{NLO EW}$
    & $\delta_\text{sum}^\text{NLO QCD}$
    & $\delta_\text{sum}^\text{NLO QCD+EW}$
    & $\delta_\text{ch.}^{\mathcal{O} (\alpha^7)}$
    & $\delta_\text{ch.}^{\mathcal{O} (\as \alpha^6)}$
    & $\delta_\text{ch.}^{\mathcal{O} (\alpha^7) + \mathcal{O} (\as \alpha^6)}$ \\
    & [\si{\percent}]
    & [\si{\percent}]
    & [\si{\percent}]
    & [\si{\percent}]
    & [\si{\percent}]
    & [\si{\percent}]
    & [\si{\percent}] \\
\midrule
             $q_\mathrm{u}\mathrm{b}$ &  82.9 &  76.3 &  59.7 &  53.1 &   -8.0 &  -28.0 &  -35.9 \\
            $\mathrm{g} q_\mathrm{u}$ &       &       &  30.6 &  30.6 \\
        $\bar{\mathrm{b}} \mathrm{g}$ &       &       &  11.1 &  11.1 \\
               $\mathrm{g}\mathrm{b}$ &       &       &  10.4 &  10.4 \\
       $\bar{q}_\mathrm{d}\mathrm{b}$ &  14.5 &  13.5 &  10.2 &   9.3 &   -6.4 &  -29.4 &  -35.8 \\
       $\bar{q}_\mathrm{d}\mathrm{g}$ &       &       &   5.5 &   5.5 \\
    $\bar{q}_\mathrm{d} q_\mathrm{u}$ &   1.9 &   1.7 &   3.0 &   2.8 &  -10.0 &   54.5 &   44.4 \\
                   $\gamma\mathrm{b}$ &       &   0.7 &       &   0.7 \\
             $\bar{\mathrm{b}}\gamma$ &       &   0.6 &       &   0.6 \\
      $\bar{\mathrm{b}} q_\mathrm{u}$ &   0.6 &   0.5 &   0.5 &   0.4 &  -13.0 &  -19.2 &  -32.1 \\
                $\gamma q_\mathrm{u}$ &       &   0.1 &       &   0.1 \\
$\bar{\mathrm{b}} \bar{q}_\mathrm{d}$ &   0.1 &   0.1 &   0.1 &   0.1 &  -11.1 &  -22.1 &  -33.3 \\
           $\bar{q}_\mathrm{d}\gamma$ &       &  0.02 &       &  0.02 \\
\bottomrule
\end{tabular}
\caption{Relative corrections for the integrated cross sections in the
  default setup for contributing channels summed over final states (column 1).
The quantities $\delta_{\text{sum}}$ in columns 2--5, defined in Eq.~\eqref{eq:channel-fractions}, give the contribution of each channel at LO, NLO EW, NLO QCD, and NLO QCD+EW relative to the sum of all channels at LO.
The quantities  $\delta_{\text{ch.}}$ in columns 6--8, defined in
Eq.~\eqref{eq:channel-corrections}, show the EW, QCD, and QCD+EW
corrections (without LO) relative to the LO of the same channel.}
\label{tab:partonic-channels}%
\end{table}%
The results given in
\refta{tab:integrated-cross-sections} for both setups are quite
similar. The relative QCD corrections amount to $31.0\%$ and $30.6\%$
and the relative EW corrections to  $-6.5\%$ and $-8.5\%$. The
difference of $2\%$ in the EW corrections caused by the additional
invariant mass cut, Eq.~\eqref{eq:invariant-mass-cut}, results from the
missing positive corrections in the radiative tail for
invariant masses $M_{\mathrm{e}^+ \mathrm{e}^-}$ below the Z~resonance
[see
\reffi{fig:mass-lepton-pair}]. The QCD scale uncertainties are reduced
by almost a factor of 2 upon including NLO QCD corrections.

In \refta{tab:partonic-channels} we list the contribution of each
partonic channel \enquote{ch.} at LO, NLO EW, NLO QCD and NLO QCD+EW,
\begin{equation}
\begin{aligned}
\delta_{\text{sum}}^{x}      &=
\frac{\sigma_{\text{ch.}}^{x}}{\sigma_{\text{sum}}^\text{LO}} \text{,} \qquad &
x = \text{LO},\ \text{NLO EW},\ \text{NLO QCD},\   \text{NLO QCD+EW} 
 \text{,}
\end{aligned}
\label{eq:channel-fractions}
\end{equation}
relative to the integrated LO cross section for the sum of all
channels in the default setup.
Entries for vanishing channels  at LO/NLO are left blank.
Furthermore, we list the size of each NLO correction (without LO) for
the specific channels relative to the LO cross section for this channel:
\begin{equation}
\delta_{\text{ch.}}^{\mathcal{O} (\alpha^7)} = \frac{\sigma^{\mathcal{O} (\alpha^7)}_{\text{ch.}}}{\sigma_{\text{ch.}}^\text{LO}} \text{,} \qquad
\delta_{\text{ch.}}^{\mathcal{O} (\alpha_\mathrm{s} \alpha^6)} = \frac{\sigma^{\mathcal{O} (\alpha_\mathrm{s} \alpha^6)}_{\text{ch.}}}{\sigma_{\text{ch.}}^\text{LO}} \text{,} \qquad
\delta_{\text{ch.}}^{\mathcal{O} (\alpha^7) + \mathcal{O} (\alpha_\mathrm{s} \alpha^6)} = \frac{\sigma_{\text{ch.}}^{\mathcal{O} (\alpha^7)} + \sigma_{\text{ch.}}^{\mathcal{O} (\alpha_\mathrm{s} \alpha^6)}}{\sigma_{\text{ch.}}^\text{LO}} \text{.}
\label{eq:channel-corrections}
\end{equation}
Since the previous quantities are only well-defined for $\sigma_{\text{ch.}}^\text{LO} \neq 0$, they are not given for channels with a gluon or photon in the initial state, for which the entry is left blank in the table.
We note that each partonic cross section when summed over all
contributing final states is IR- and collinear safe,
but unphysical in the sense that it can not be measured.
Nevertheless, we find the results in \refta{tab:partonic-channels} useful to trace back some of the effects we see in the differential distributions given below.


While the dominant partonic channel, $q_\Pu\Pb$, makes up $83\%$
of LO, its contribution is reduced to $53\%$ at NLO, mainly owing to
the appearance of gluon-induced channels at NLO QCD (see Table
\ref{tab:partonic-channels}), which make up almost half of the NLO cross section.
The anti-top production channels with the initial states
$\bar{\mathrm{b}}\mathrm{g}$ (with sample Feynman diagram
shown in \reffi{fig:antitopdiag}) are with $11\%$ the third-most important
contribution for the integrated cross section at NLO QCD; these channels will become important for the discussion of the differential distributions.
At NLO EW, the corresponding channels with the
$\bar{\mathrm{b}}\gamma$ initial states only contribute $0.6\%$.
Channels that are non-zero at LO have typically negative corrections
that are larger than those for the cross section summed over all
partonic channels. These negative corrections are, however, partially
compensated by additional channels that are non-zero only at NLO.  The
only individual channel receiving positive QCD corrections is the one with the
$\bar{q}_\mathrm{d} q_\mathrm{u}$ initial state, which only receives
$s$-channel contributions, while all other partonic channels receive only
$t$-channels contributions at LO.
The relative EW corrections to the individual channels range between $-6\%$ and
$-13\%$.

\subsection{Comparison with literature results}
\label{sec:comparison-with-literature-results}

\begin{table}
\centering
\begin{tabular}{l@{\hspace{2em}}S[table-format=1.2]@{\hspace{2em}}S[table-format=1.3]S[table-format=1.3]}
\toprule
& {Ref.~\cite{Pagani:2020mov}} & \multicolumn{2}{c}{Z-peak setup without $\bar\Pb\Pg/\bar\Pb\gamma$ channels} \\
& & \multicolumn{1}{c}{w/o decay corr.} & \multicolumn{1}{c}{w/ decay corr.} \\
\midrule
 NLO QCD/LO          & 1.24 & 1.289 & 1.195 \\
(NLO QCD+EW)/NLO QCD & 0.93 & 0.919 & 0.924 \\
\bottomrule
\end{tabular}
\caption{Comparison with results denoted as $\mathrm{t}\ell^+\ell^-\mathrm{j}$ (Z-peak, 5FS) in Table~2 of Ref.~\cite{Pagani:2020mov}.
The corrections to the semi-leptonic top decay are subtracted from our
predictions in  column 3 but not in column 4. Contributions of the $\bar\Pb\Pg$ and  $\bar\Pb\gamma$ channels are
omitted in columns 3 and 4.}
\label{tab:comparison-with-pagani-et-al}
\end{table}
In \refta{tab:comparison-with-pagani-et-al} we compare
\begin{enumerate}
\item the result from Table~2 of \citere{Pagani:2020mov} for the
  five-flavour scheme (5FS) including $t$, $s$ and
  $\mathrm{t}\mathrm{W}_\mathrm{h}$\footnote{Following \citere{Pagani:2020mov}, $\mathrm{t}\mathrm{W}_\mathrm{h}$ denotes the production of an on-shell top quark (with leptonically decaying Z~boson) and a $\mathrm{W}$~boson decaying into a quark pair.
These contributions naturally arise at NLO EW in both the on-shell and fully off-shell calculation.} channels for the Z-peak region, with
\item our results in the Z-peak setup, where the additional cut Eq.~\refeq{eq:invariant-mass-cut}
approximately implements $| M_{\ell^+\ell^-} - M_\mathrm{Z} | < \SI{10}{\giga\electronvolt}$ of \citere{Pagani:2020mov}.
Furthermore we exclude the initial states $\bar{\mathrm{b}}
\mathrm{g}$ and $\bar{\mathrm{b}} \gamma$.
These channels would correspond to $\bar{\mathrm{t}} \mathrm{Z}\mathrm{W}^+$ production in an on-shell approximation and are not included in \citere{Pagani:2020mov}.
Finally, we subtract the relative NLO QCD and NLO EW corrections of
the leptonic top decays, which are included in our calculation for
off-shell tops but not in the on-shell calculation of
\citere{Pagani:2020mov}. Specifically we approximate
\begin{align}
\left.\frac{\sigma^\text{NLO QCD}}{\sigma^\text{LO}}\right|_{\text{on
    shell}} \approx{}&
\frac{\sigma^\text{NLO QCD}}{\sigma^\text{LO}} - \delta^{\alpha_\mathrm{s}} \text{,} \notag\\
\left.\frac{\sigma^\text{NLO QCD+EW}}{\sigma^\text{LO}}\right|_{\text{on
    shell}} \approx{}&
\frac{\sigma^\text{NLO QCD+EW} - (\delta^{\alpha_\mathrm{s}} + \delta^{\alpha}) \sigma^\text{LO}}{\sigma^\text{NLO QCD} - \delta^{\alpha_\mathrm{s}} \sigma^\text{LO}} \text{,}
\end{align}
with $\delta^{\alpha_\mathrm{s}} = \SI{-9.38}{\percent}$ and
$\delta^{\alpha} = \SI{1.34}{\percent}$ taken from Table~1 of \citere{Basso:2015gca}.
For comparison, we also give our predictions which include top-quark-decay corrections
 (last column).
\end{enumerate}
Both setups are similar in terms of phase-space cuts. Important
differences between the two setups are that we include the full
off-shell effects but omit the charge-conjugated
$\bar{\mathrm{t}}\mathrm{Z}\mathrm{j}$ process.
However, for a comparison of relative corrections we do not expect
this to make a large difference. The contribution of
$\bar{\mathrm{t}}\mathrm{Z}\mathrm{j}$ production is only  roughly
$1/3$ (see Table~3 of \citere{Campbell:2013yla}). As can be seen in
\refta{tab:comparison-with-pagani-et-al}, we find good agreement for
the relative EW corrections within about one per cent. The difference
in the relative QCD corrections of $5\%$ seems reasonable in view of
the described differences of the calculations.
It is worth noticing that the QCD scale uncertainties reported in
Table~1 of \citere{Pagani:2020mov} for the 5FS scheme are smaller than those
we obtain (see Table~\ref{tab:integrated-cross-sections}). This is motivated
by a slightly different central-scale choice which seems to underestimate
artificially the QCD uncertainties already at LO, making it important to
compare 5FS and 4FS results. The  QCD scale uncertainties found in our
5FS setup are almost as large as the combined   5FS and 4FS scale
uncertainties advocated in  \citere{Pagani:2020mov}.

Finally, we compare our results for the NLO QCD corrections relative
to the LO with those of \citere{Campbell:2013yla}.  Combining the
results of the  \enquote{Standard cuts} setup with 2 and 3 jets given in
 Table~2 of \citere{Campbell:2013yla} we find
\begin{equation}
\delta = \frac{0.585 + 0.693}{1.05} = \SI{121.7}{\percent} \text{.}
\end{equation}
Subtracting the contributions of the $\bar\Pb\Pg$ partonic channels involving anti-top
production given in \refta{tab:partonic-channels} from our result in
the default setup in
\refta{tab:integrated-cross-sections} (relative NLO QCD) yields
\begin{equation}
\delta = \SI{131.0}{\percent} - \SI{11.1}{\percent} = \SI{119.9}{\percent} \text{.}
\label{eq:nlo-qcd-no-anti-top}
\end{equation}
The two numbers agree within $2\%$, which we find acceptable in view of
the differences between both calculations.
Again we use a relative correction to minimize the sensitivity due to differences in the setup of \citere{Campbell:2013yla}, which uses a hadronic centre-of-mass energy of \SI{14}{\tera\electronvolt}, includes the $\bar{\mathrm{t}}\mathrm{Z}\mathrm{j}$ contributions and has slightly different cuts.
However, it does include corrections to the top-quark decay, as do the numbers in Eq.~\eqref{eq:nlo-qcd-no-anti-top}.

\subsection{Differential distributions}
\label{sec:differential-distributions}

We now discuss the differential distributions presented in Figs.~\ref{fig:rap-jets}--\ref{fig:miscellaneous}.
Each figure contains three panels, showing 1) absolute predictions for
the LO, NLO QCD and NLO QCD+EW,  
2) the relative corrections of NLO QCD,
 and 3) NLO EW normalised to the LO.
Each panel also contains NLO predictions without the anti-top
production channels, Eq.~\eqref{eq:anti-top-production}.
For LO and NLO QCD predictions
uncertainty bands are included, estimating the
missing higher QCD orders using the envelope from a 7-point scale
variation, Eq.~\eqref{eq:scale-variations}, of the cross sections. 

\subsubsection{Jet observables}

\begin{figure}
\centering
\subfigure[\label{fig:rap-top-jet}]{\includegraphics[width=0.49\textwidth]{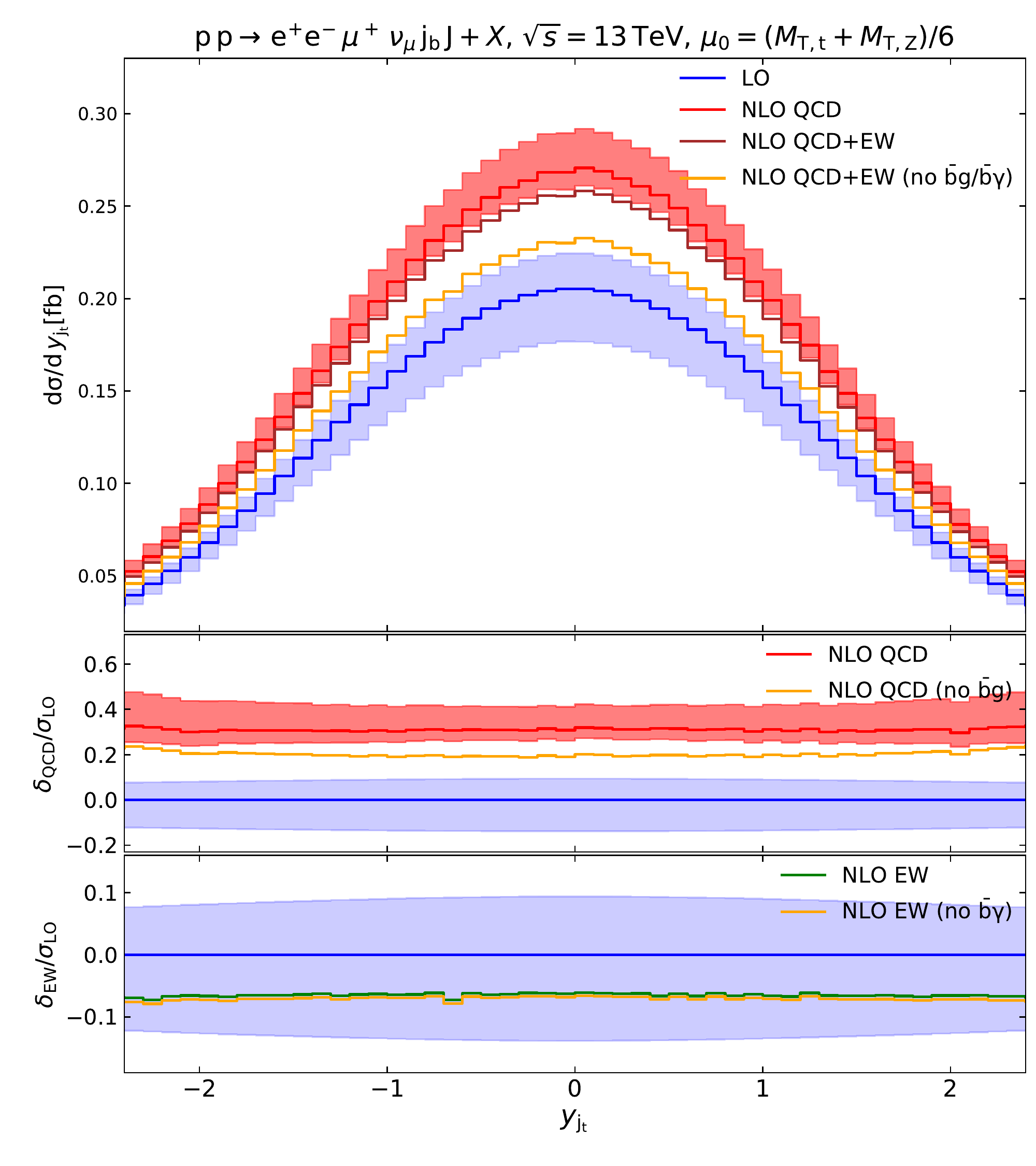}}\hspace{0.01\textwidth}%
\subfigure[\label{fig:rap-spectator-jet}]{\includegraphics[width=0.49\textwidth]{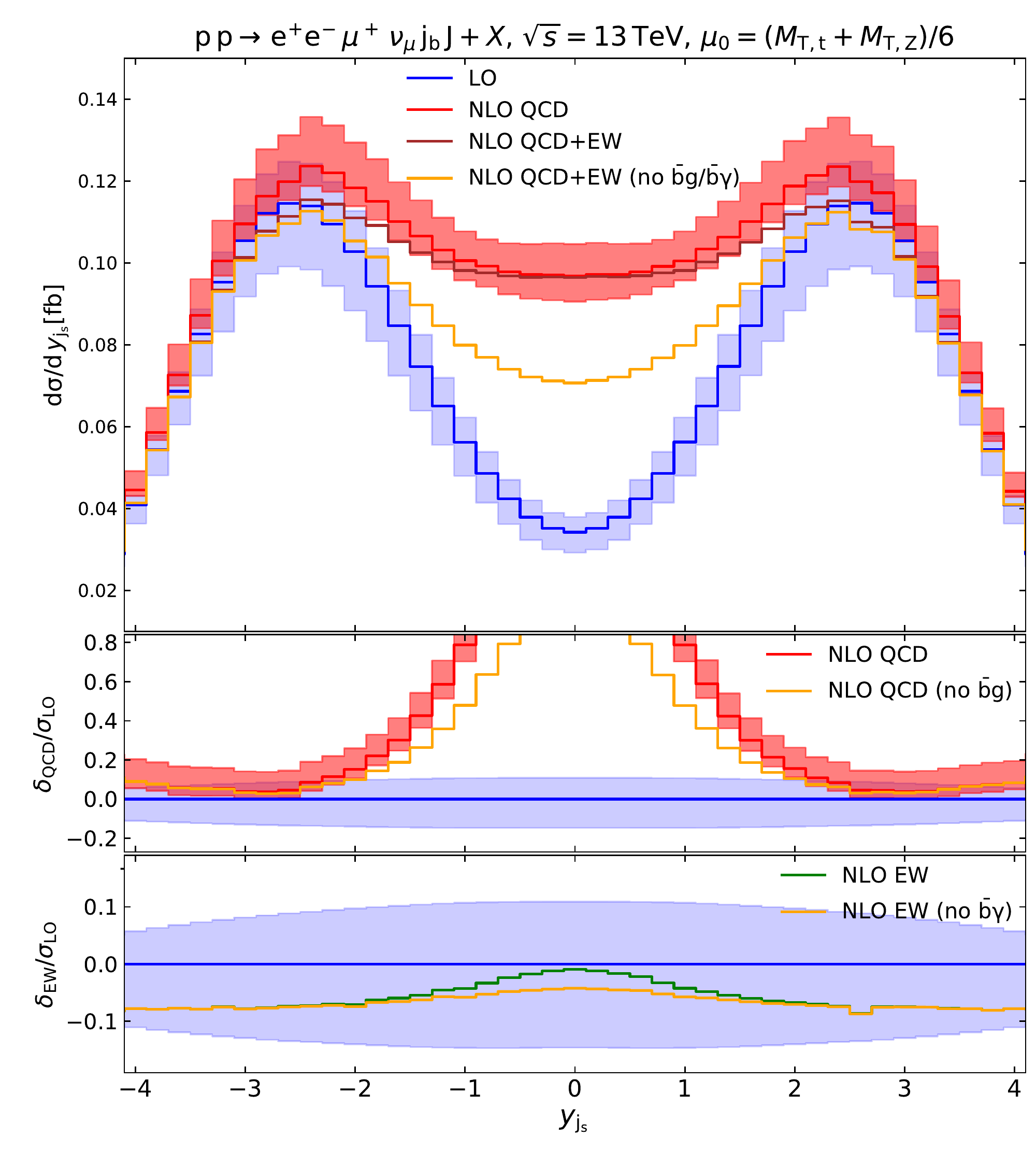}}
\caption{Rapidity distributions for the top-decay jet $\mathrm{j}_\mathrm{b}$ and spectator jet $\mathrm{j}_s$.
See \refse{sec:kinreco} for the definition of these objects.}
\label{fig:rap-jets}
\end{figure}
Figure~\ref{fig:rap-top-jet} presents the cross section differentially
in the rapidity of the top-decay jet and
\reffi{fig:rap-spectator-jet} in the rapidity of the spectator jet, as defined in \refse{sec:kinreco}.
The top-decay jet is mainly produced in the central region, $|y| < 2$,
whereas the behaviour of the spectator is determined by the
$t$-channel $\PW$-boson propagator attached to the spectator quark
line [see for example \reffi{fig:lo_t}] and therefore preferably produced in the
forward region, peaking at $|y| \approx 2.5$, similar to the tagging
jets in vector-boson scattering.
At NLO QCD the two distributions are differently affected by the
corrections: while the relative corrections are flat for the top-decay jet,
they fill up the central region for the spectator jet, with a large
contribution from the $\bar\Pb\Pg$ channels and an almost as large
contribution from the $\Pb\Pg$ channels.%
\footnote{The $\Pb\Pg$ channels correspond to the
  $\mathrm{t}\mathrm{W}_\mathrm{h}$-channel contribution of 
\citere{Pagani:2020mov}, where a similar effect has been observed.}
Extra gluon radiation from the spectator jet changes its direction filling
effectively the central region,  which is disfavoured at LO for the spectator jet.
Moreover, the event reconstruction (see \refse{sec:kinreco})  enables to tag the gluon as the
spectator jet, giving a rapidity spectrum which is in general more central
than the one of the quark orginated from the $t$-channel topology.
The EW corrections for the top-decay-jet rapidity distribution are
practically constant and reproduce those of the integrated cross section.
For the spectator jet, we observe a similar picture in the dominant region $|y| \gtrsim 2$, but
smaller EW corrections in the central region in accordance with the
results of the on-shell calculation \cite{Pagani:2020mov}. 
The contribution of the $\bar\Pb\gamma$ channels tends to cancel the
EW corrections near $y_{\sj} \approx 0$.

\begin{figure}
\centering
\subfigure[\label{fig:pt-top-jet}]{\includegraphics[width=0.49\textwidth]{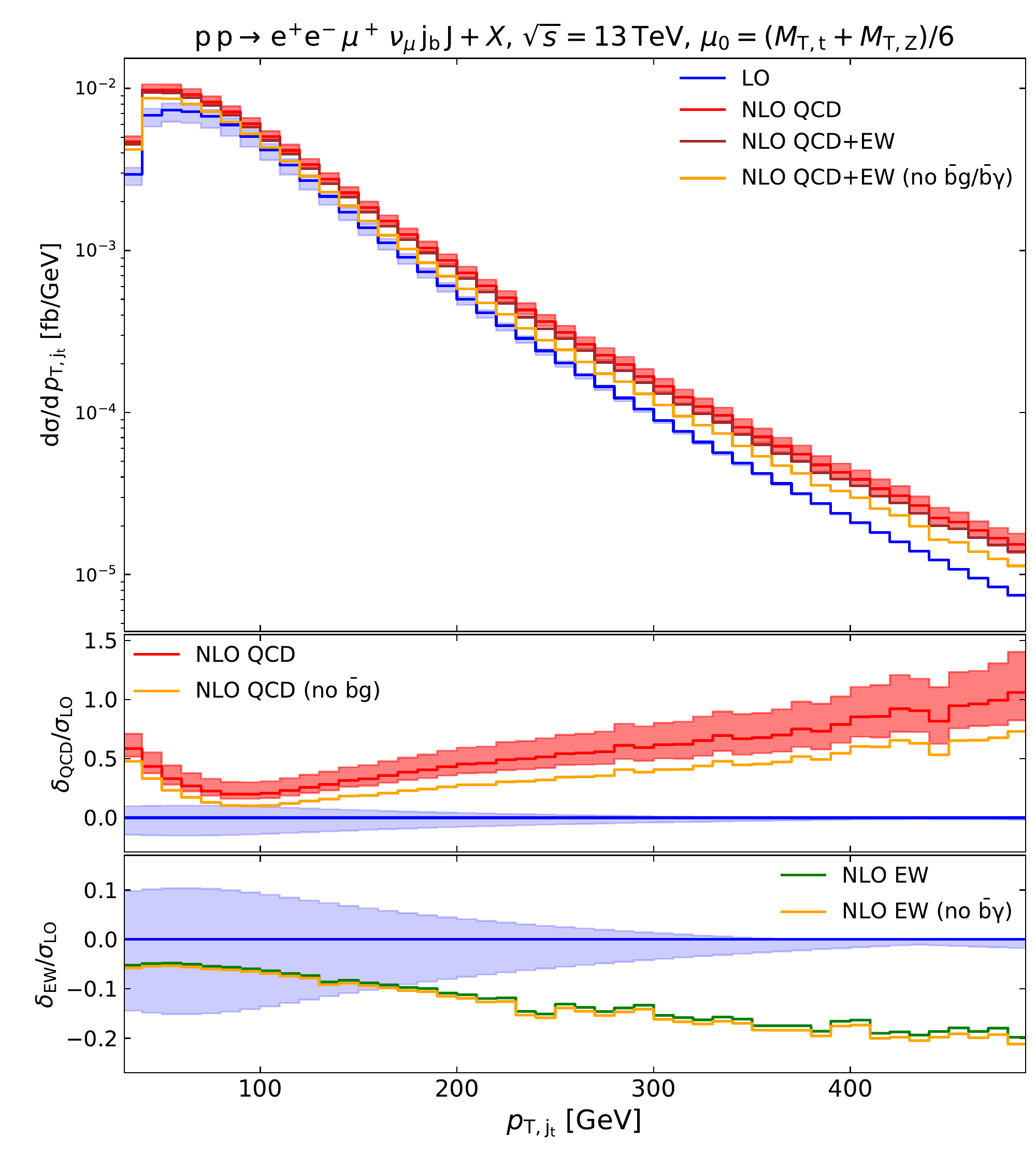}}\hspace{0.01\textwidth}%
\subfigure[\label{fig:pt-spectator-jet}]{\includegraphics[width=0.49\textwidth]{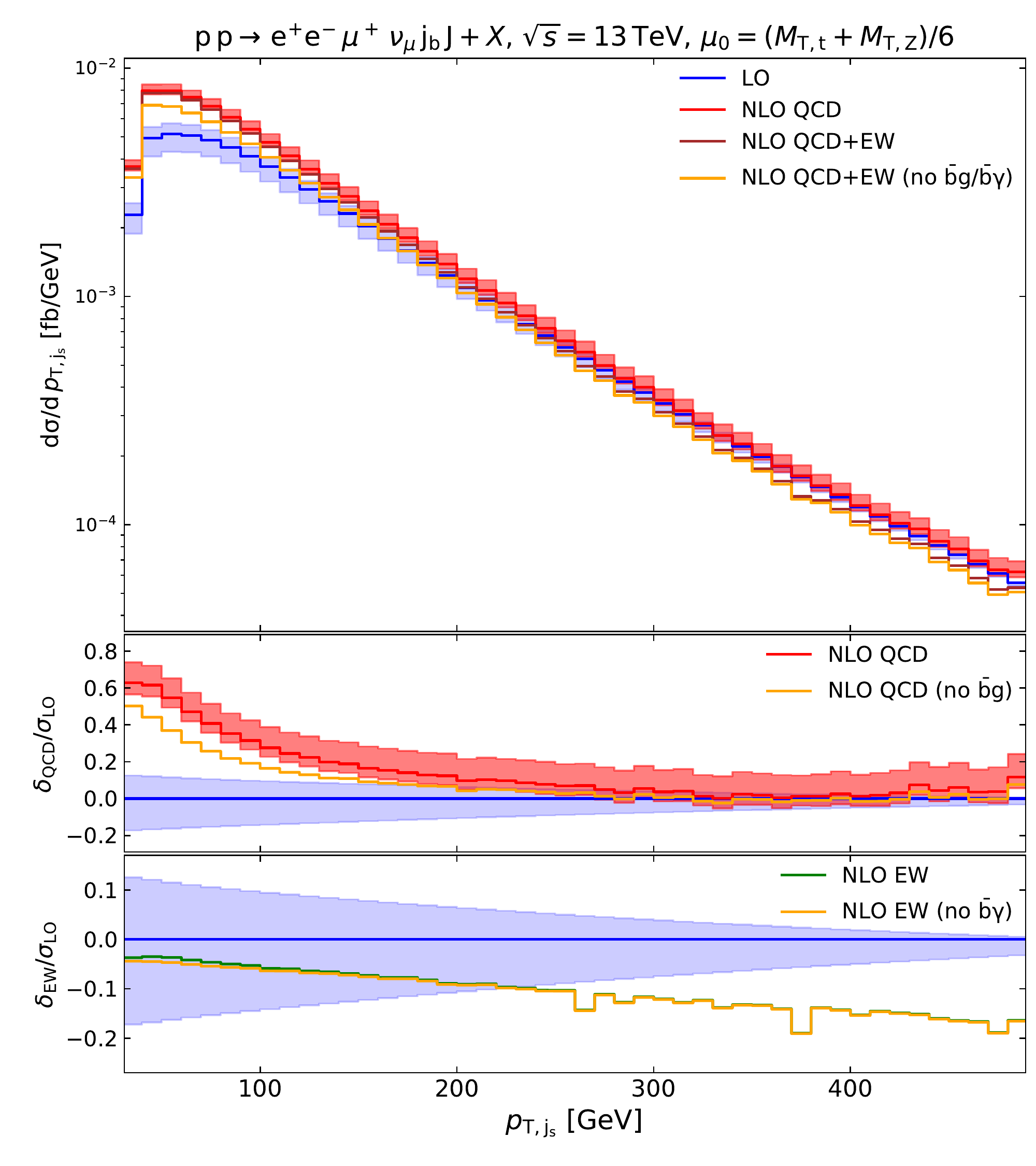}}
\caption{Transverse-momentum distributions for the top-decay jet $\mathrm{j}_\mathrm{b}$ and spectator jet $\mathrm{j}_s$.}
\label{fig:pt-jets}
\end{figure}
The distributions of the transverse momenta of the top-decay jet
and the spectator jet,%
\footnote{The reduction of the LO scale dependence for high values of
  the transverse momenta has already been observed in
  \citere{Demartin:2015uha} for a similar scale choice and suspected
  to be an artefact of the scheme.} presented in
\reffis{fig:pt-top-jet}--\ref{fig:pt-spectator-jet}, both receive
\SIrange{25}{60}{\percent} QCD corrections in the region $p_{\rT} <
\SI{100}{\GeV}$.  However, in the large-$p_{\rT}$ regions the QCD
corrections to the spectator-jet distribution approach zero, while
those for the top-decay jet increase up to \SI{100}{\percent}.  The
striking difference between the two observables is due to the different
LO behaviour, where the top-decay jet (resulting from the massive top
quark) is much softer than the
spectator jet (produced directly), while at NLO QCD the opening of the partonic channels
with a gluon and a light quark ($\Pg q$) 
in the initial state enhances 
the tail of both distributions.  The LO behaviour is very similar to
the one found in $t$--channel single-top production
\cite{Frederix:2019ubd}, where the spectator recoils against the top
quark, and therefore the transverse momentum of the
top quark (balancing the one of the spectator jet) is shared among
its decay products, resulting in a top-decay jet with smaller
$p_{\rT}$ than the spectator jet. Since in $\Pt\PZ\Pj$ production the
additional $\PZ$ boson tends to be soft or closer in phase space to the top
quark than to the spectator jet, the same argument
applies explaining the relative softness of the top-decay jet.  For both distributions in \reffi{fig:pt-jets} the
EW corrections grow negatively from roughly
\SIrange{-5}{-20}{\percent} in the large-$p_\mathrm{T}$ region.  The
relative EW corrections to the top-decay-jet transverse-momentum distribution are similar
to those for the reconstructed top quark, showing in the tail
the same Sudakov behaviour as in inclusive calculations
(see Fig.~6 of \citere{Pagani:2020mov}). Our results for the relative EW corrections to
the distribution in the transverse momentum of the spectator jet agree qualitatively
with those for the transverse momentum of the leading light jet in
Fig.~6 of \citere{Pagani:2020mov}.

\begin{figure}
\centering
\subfigure[\label{fig:mass-jets}]{\includegraphics[width=0.49\textwidth]{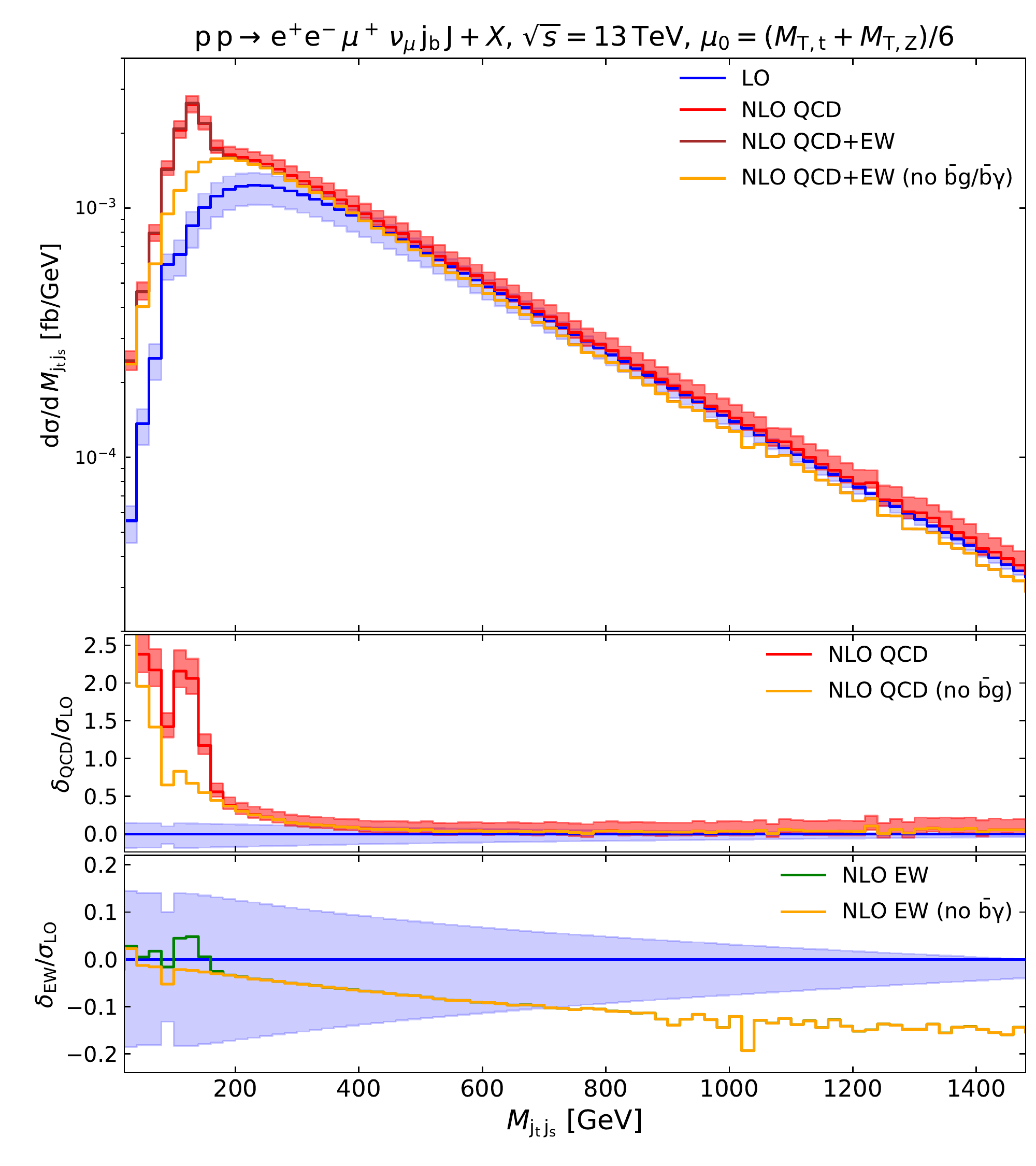}}\hspace{0.01\textwidth}%
\subfigure[\label{fig:rap-sep-jets}]{\includegraphics[width=0.49\textwidth]{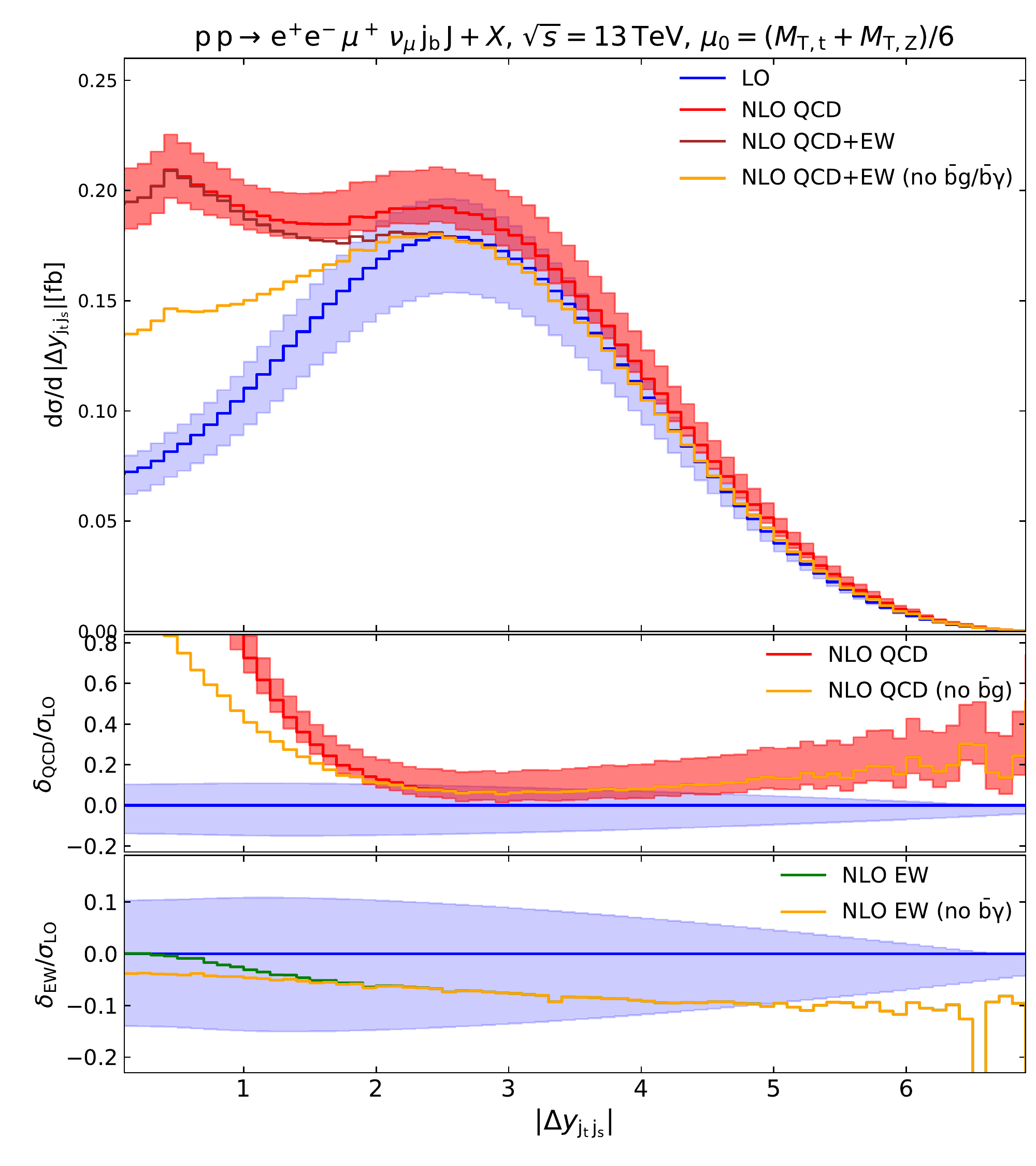}}
\caption{Invariant-mass and rapidity-separation distributions of the
  jet pair.}
\label{fig:dijet-observables}
\end{figure}
Looking at observables involving both the top-decay and spectator jet, we find a resonance in the invariant-mass distribution in \reffi{fig:mass-jets} at around $M_{\mathrm{j}_\mathrm{t} \mathrm{j}_s} \approx \SI{165}{\giga\electronvolt}$.
This resonance comes from an anti-top quark, which is produced
starting from NLO QCD and EW in processes of the type given in
Eq.~\eqref{eq:anti-top-production} and decays
hadronically, for example $\bar{\mathrm{t}} \to \bar{\mathrm{b}}
\bar{\mathrm{u}} \mathrm{d}$ [see \reffi{fig:antitopdiag}]. This
contribution is clearly visible in the relative QCD and EW corrections.
Since the two-jet invariant mass does not capture all three quarks of
the anti-top decay---only when two of them are clustered
together---the peak is below the top-quark mass.
The EW corrections increase negatively up to \SI{-15}{\percent} for invariant masses
of \SI{1500}{\giga\electronvolt}. The QCD corrections
diminish from $+25\%$ to $+10\%$ above $200\GeV$.

In the distribution of the rapidity separation of the two jets,
\reffi{fig:rap-sep-jets}, we observe a rapidity gap of $|\Delta
y_{\mathrm{j}_\mathrm{t} \mathrm{j}_s}| \approx 2.5$ at LO, as
expected from the rapidities of the individual jets [see \reffis{fig:rap-top-jet} and \ref{fig:rap-spectator-jet}].
At NLO QCD this gap is filled, which is consistent with the
observation that strong corrections mostly affect the central regions
of each jet distribution, as seen in rapidity spectra of individual jets [see \reffi{fig:rap-top-jet}].
A large fraction of events for small $|\Delta y_{\mathrm{j}_\mathrm{t}
  \mathrm{j}_s}|$ originates from the $\bar\Pb\Pg$ anti-top production
channels and the $\Pb\Pg$ channels at NLO QCD.
The corresponding $\bar\Pb\gamma$ channels cancel the negative NLO EW
corrections in the rapidity gap, which grow up to \SI{-12}{\percent} for $|\Delta y_{\mathrm{j}_\mathrm{t} \mathrm{j}_s}| \approx 6$.

\subsubsection{Leptonic observables}

\begin{figure}
\centering
\subfigure[\label{fig:mass-lepton-pair}]{\includegraphics[width=0.49\textwidth]{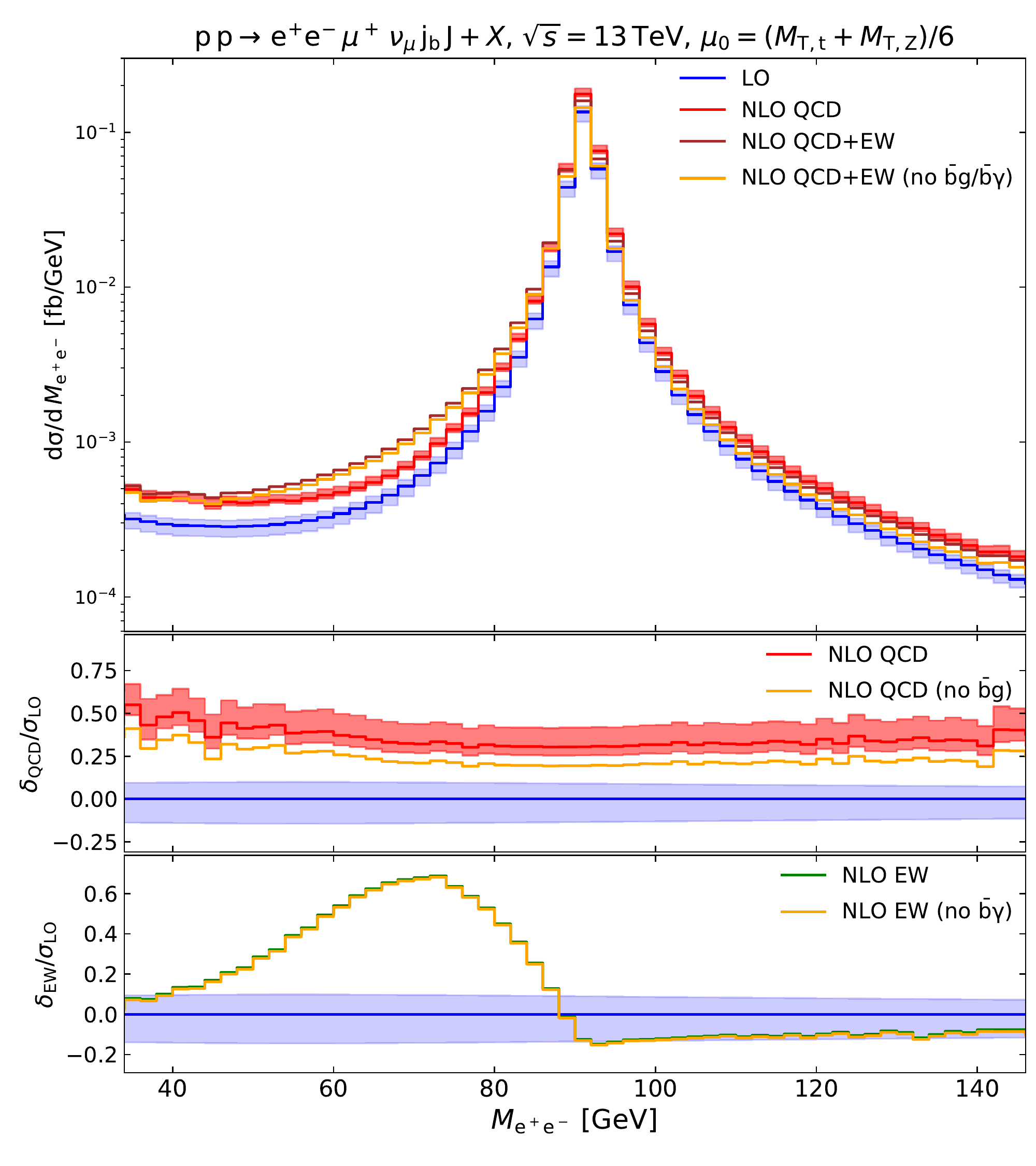}}\hspace{0.01\textwidth}%
\subfigure[\label{fig:mass-three-leptons}]{\includegraphics[width=0.49\textwidth]{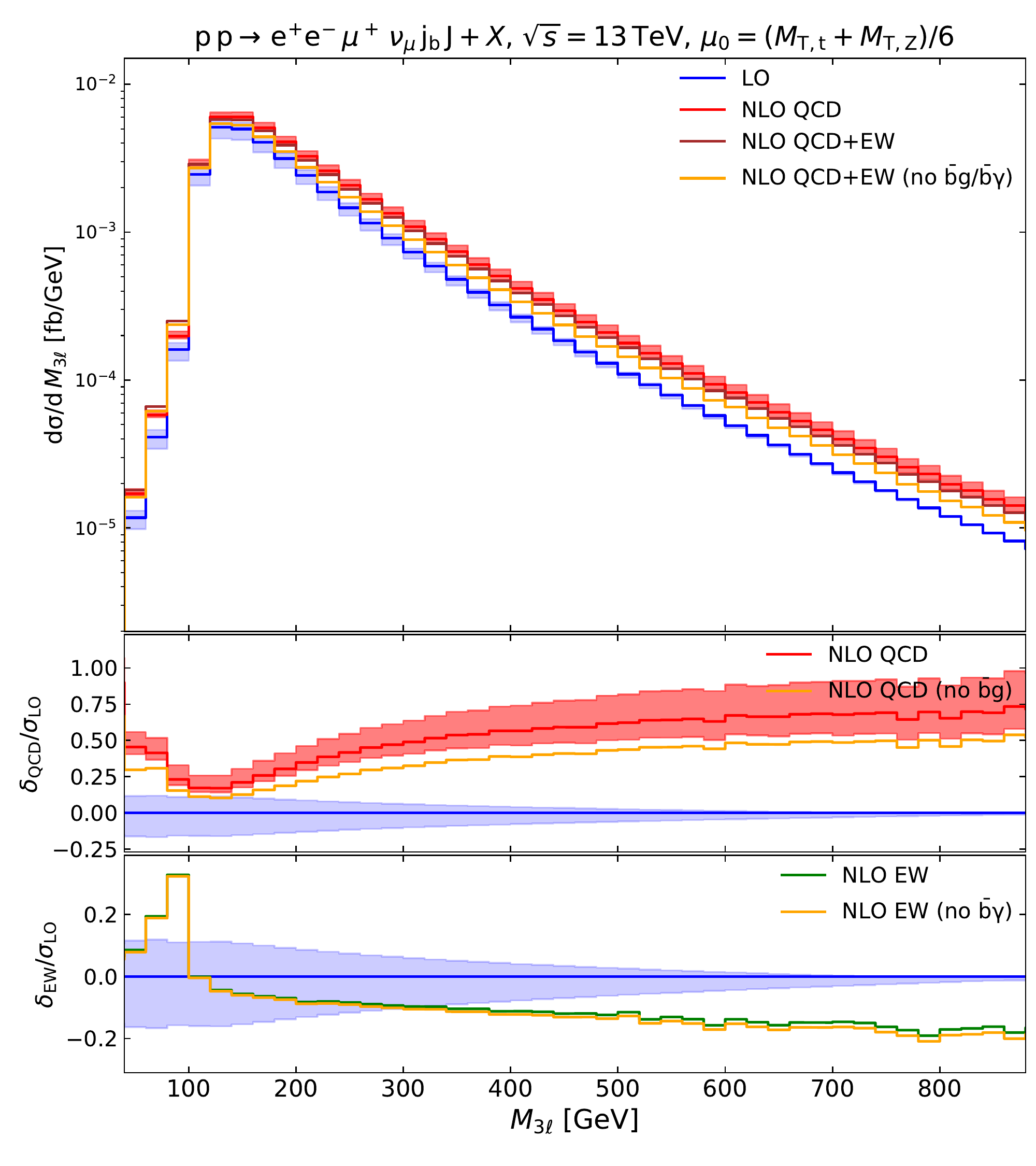}}
\caption{Invariant-mass distributions of the same-flavour lepton pair and of all three charged leptons.}
\label{fig:masses-leptons}
\end{figure}
Coming to leptonic observables, we see the Breit--Wigner shape from the Z-boson decay around its mass in the $M_{\mathrm{e}^+ \mathrm{e}^-}$ distribution in \reffi{fig:mass-lepton-pair}.
At NLO EW, soft-photon radiation causes positive corrections of up to \SI{60}{\percent} below the Z-boson mass around \SI{70}{\giga\electronvolt}.
This radiative return is very similar in terms of size and shape to Drell--Yan lepton-pair production.
The NLO QCD corrections around the resonance  are flat and reproduce those of the integrated cross section.

A similar effect of NLO EW corrections arises in the $M_{3\ell}$ observable in
\reffi{fig:mass-three-leptons}. The region below the peak is filled by
events where a real photon emitted from one of the decay leptons is not
included in $M_{3\ell}$, resulting in corrections of up to \SI{35}{\percent}.
Negative NLO EW corrections of up to \SI{-20}{\percent} arise for very
large invariant masses.
The NLO QCD corrections reach $70\%$ for small and large $M_{3\ell}$,
while they are about $20\%$ near the peak of the distribution.

\begin{figure}
\centering
\subfigure[\label{fig:pt-anti-muon}]{\includegraphics[width=0.49\textwidth]{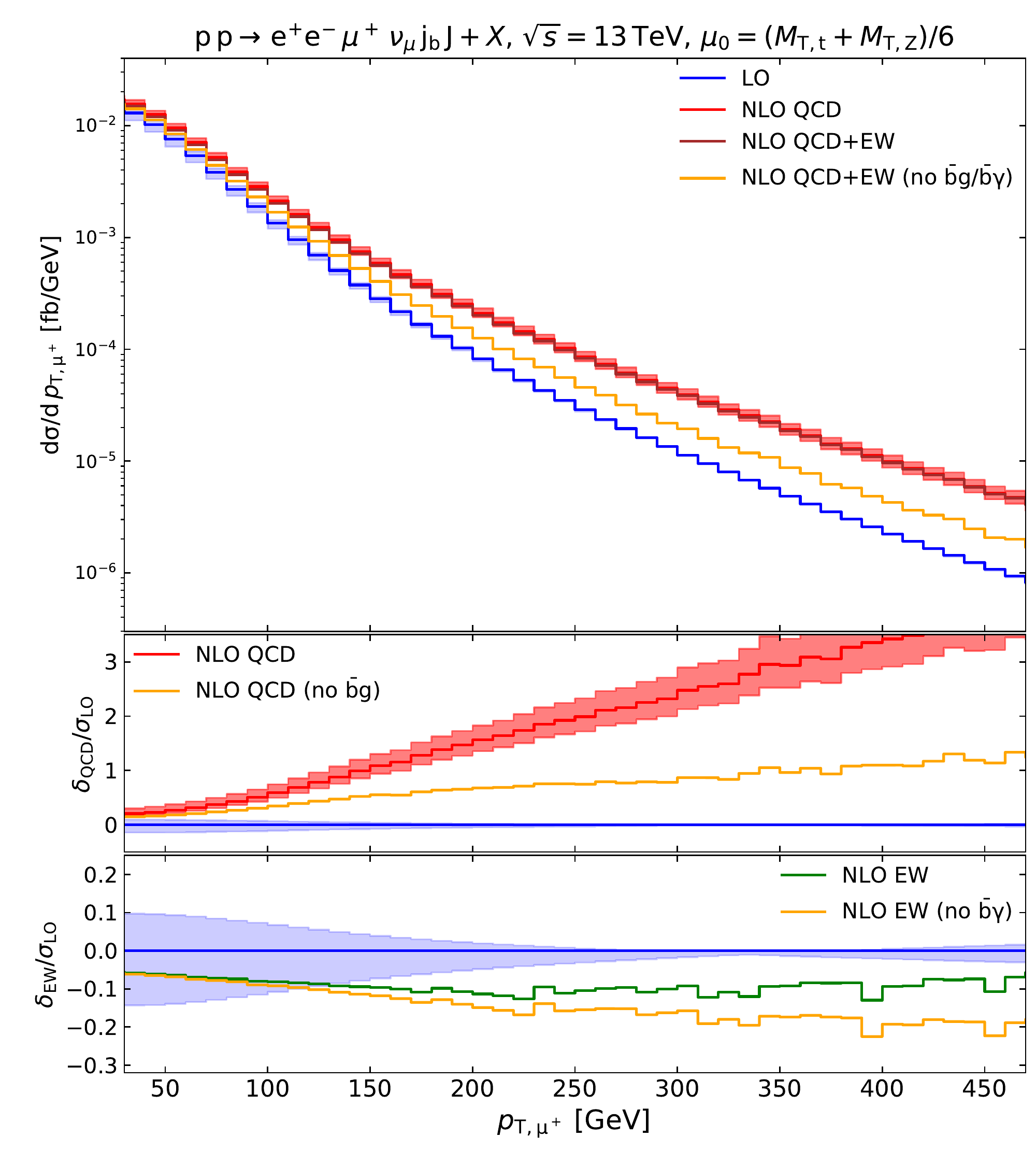}}\hspace{0.01\textwidth}%
\subfigure[\label{fig:pt-electron}]{\includegraphics[width=0.49\textwidth]{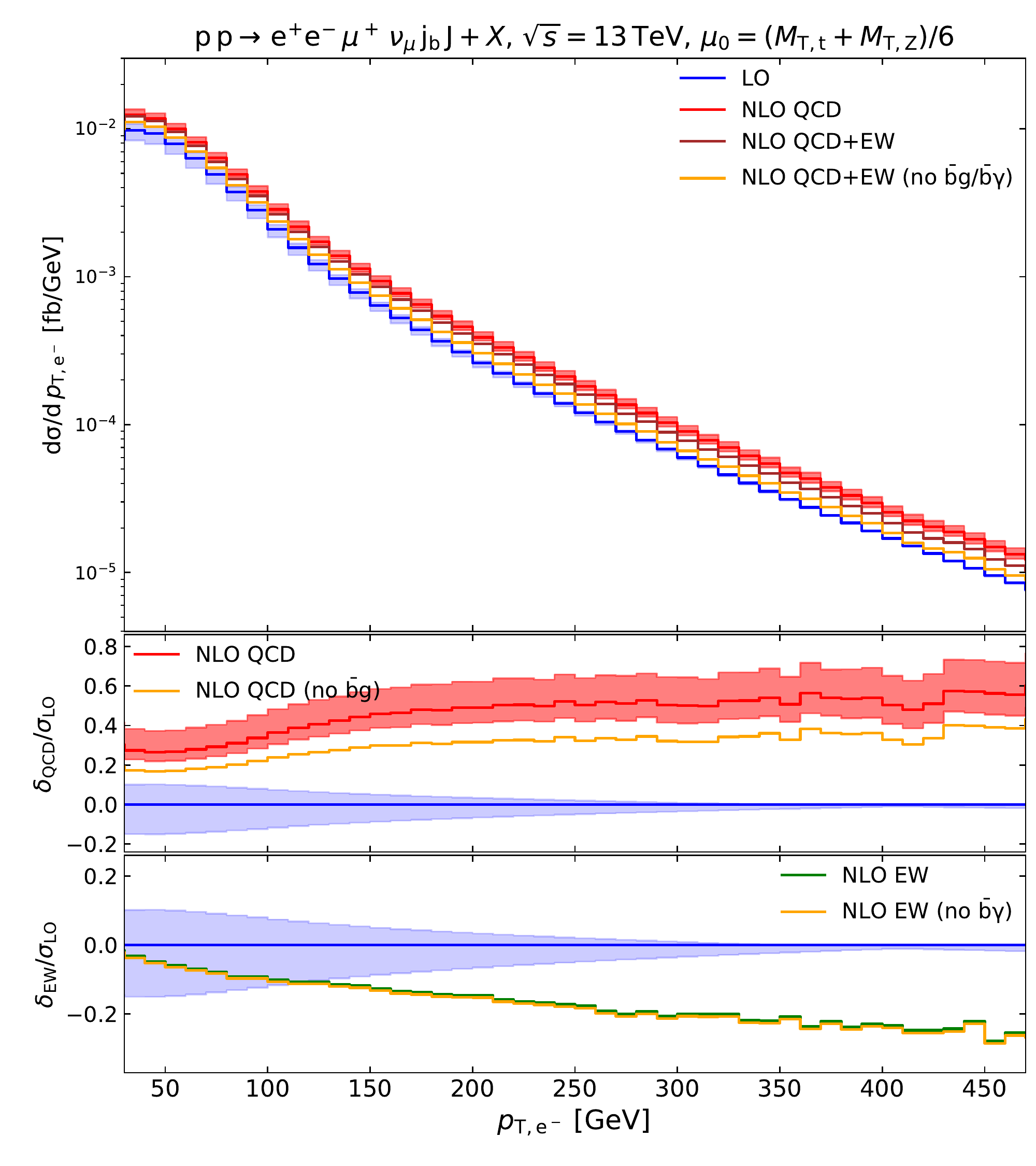}}
\caption{Transverse-momentum distributions of the anti-muon and the electron.}
\label{fig:pt-momenta-leptons}
\end{figure}
In \reffi{fig:pt-momenta-leptons} we show transverse-momentum
distributions for the anti-muon and the electron.
The EW corrections are comparably flat for the anti-muon due to
the positive contribution of the $\bar\Pb\gamma$ channels, cancelling most
of the negative NLO EW corrections for large transverse momenta.
For the electron transverse momentum, the NLO EW corrections grow negatively up to
\SI{-20}{\percent} around $500\GeV$, suggesting the dominance of
EW Sudakov logarithms.
It is worth noticing that the anti-muon is a product of the decay of
the top~quark, while the electron comes from the decay of a $\PZ$ boson,
suggesting different behaviours in the enhancement of
EW Sudakov logarithms of virtual origin.
For what concerns QCD corrections, very different behaviours are found
for the two charged leptons. The relative QCD corrections
to the electron transverse-momentum distribution increase
from $25\%$ to $50\%$ in the moderate-$p_{\rT}$ region and
flatten out for values larger than $200\GeV$. The transverse-momentum
distribution of the anti-muon is much stronger affected by QCD corrections, which enhance
the LO distribution by up to $400\%$. 
This huge effect can be explained with polarisation arguments.
At LO the $\PW^+$~bosons (decaying into $\mu^+\nu_\mu$) mostly come from the
top-quark decay and are therefore either longitudinal (dominant) or left handed
(subdominant). As a consequence the anti-muons resulting from the
$\PW$ decays are preferably emitted in the direction opposite
to the high-energetic $\PW^+$ boson and therefore have softer $p_{\rT}$.
At NLO QCD the large anti-top contribution in the $\bar{\Pb}\Pg$ channel
is characterised by $\PW^+$ bosons
that are preferably right handed (as they originate from the anti-top quark
via a helicity-right-handed coupling) leading to emitted anti-muons
collinear to the $\PW^+$ bosons and therefore enhancing the high-$p_{\rT}$ tail.

\begin{figure}
\centering
\subfigure[\label{fig:transverse-mass}]{\includegraphics[width=0.49\textwidth]{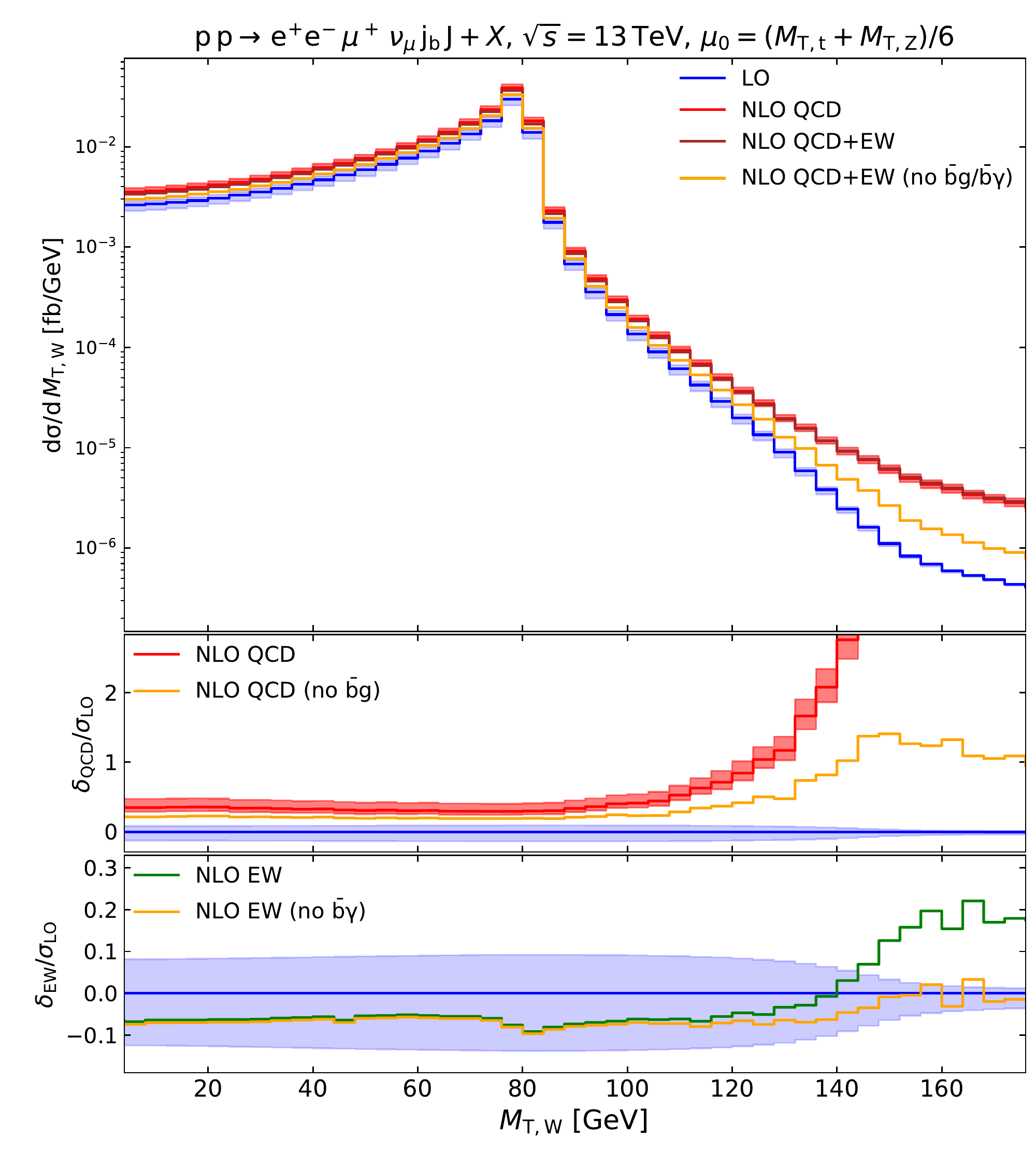}}\hspace{0.01\textwidth}%
\subfigure[\label{fig:azimuthal-angle-lepton-pair}]{\includegraphics[width=0.49\textwidth]{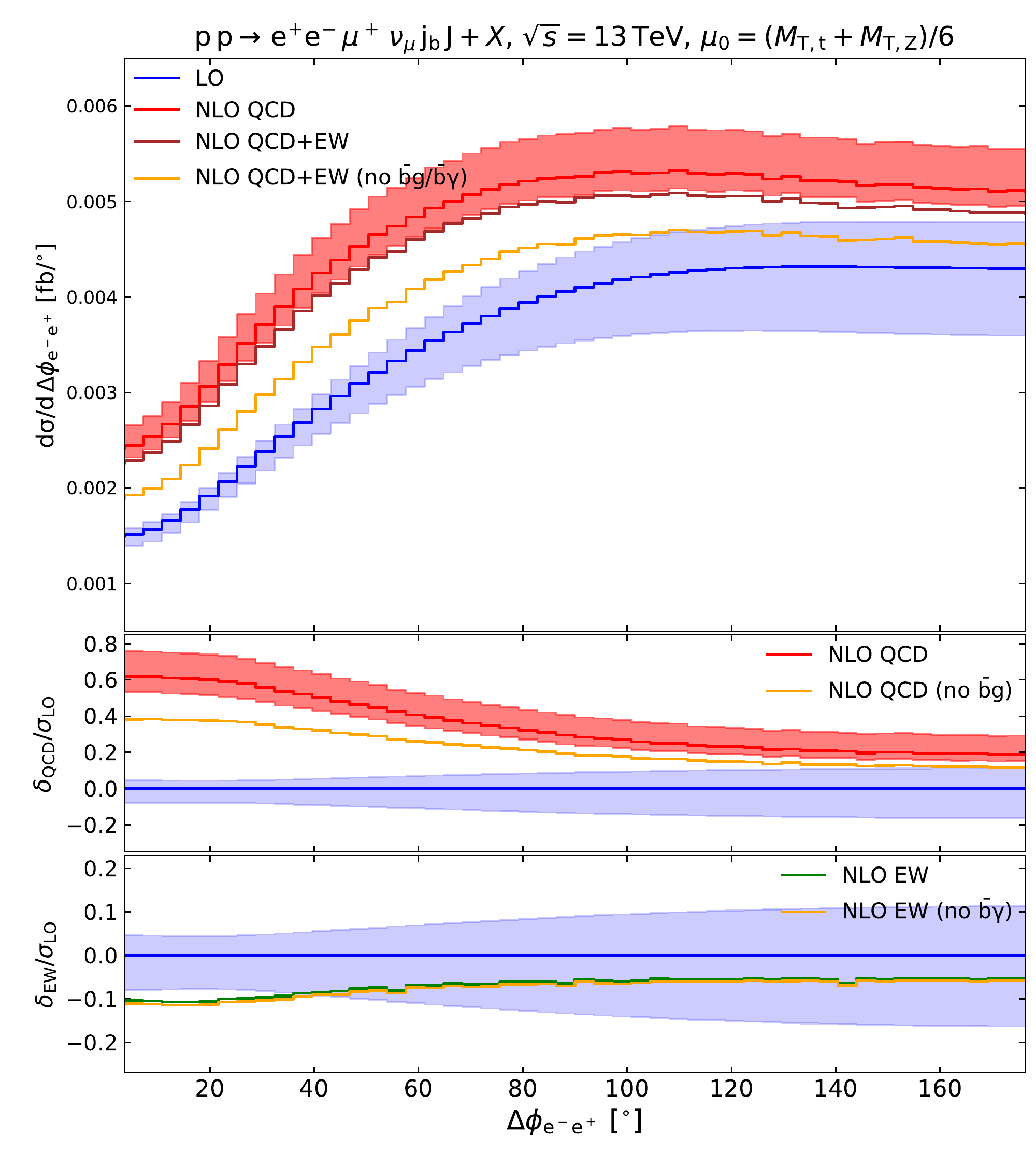}}
\caption{Transverse-mass distribution of the W boson as defined in Eq.~\eqref{eq:w-transverse-mass} and azimuthal-angle-separation distribution of the $\Pe^+\Pe^-$ pair.}
\label{fig:angles}
\end{figure}
In \reffi{fig:transverse-mass} we present the distribution in the transverse mass of the W boson, defined as
\begin{equation}
M_\mathrm{T}^\mathrm{W} = \sqrt{2 p_\mathrm{T}^{\mu^+} p_{\mathrm{T},\text{miss}} (1 - \cos \Delta \phi_{\mu^+ \text{miss}})} \text{,}
\label{eq:w-transverse-mass}
\end{equation}
where $p_{\mathrm{T},\text{miss}}$ is the missing transverse momentum
and $\Delta \phi_{\mu^+ \text{miss}}$ the azimuthal angle between the
transverse momentum of the anti-muon and the missing transverse momentum.
This distribution is very similar to the one measured in the
charge-current Drell--Yan process with the characteristic edge at around $M_\mathrm{W}$.
The NLO QCD and EW corrections are flat in a region up to
\SI{100}{\giga\electronvolt}, after which they increase mainly due to the anti-top production channels
$\bar\Pb\Pg/\bar\Pb\gamma$, in which the charged-lepton--neutrino pair is not
produced in a top-quark decay. While contributions with a
(transverse) mass of the W~boson above the top-quark mass
cannot contain a resonant top quark, they can still contain a resonant
anti-top quark.

Figure \ref{fig:azimuthal-angle-lepton-pair} displays the cross section differentially in the azimuthal angle between the $\Pe^+\Pe^-$ pair.
At LO the electron and positron are more likely to be back-to-back ($\Delta \phi_{\mathrm{e}^- \mathrm{e}^+} > \SI{90}{\degree}$) than along each other ($\Delta \phi_{\mathrm{e}^-\mathrm{e}^+} < \SI{90}{\degree}$).
The NLO EW corrections vary from \SIrange{-10}{-5}{\percent}.
At NLO QCD, the corrections are larger for lepton pairs close to each
other in the azimuthal plane, which typically result from
high-energetic $\PZ$ bosons, and smaller for the back-to-back configuration.

\subsubsection{Lepton--jet observables}

\begin{figure}
\centering
\subfigure[\label{fig:mass-top-jet-anti-muon}]{\includegraphics[width=0.49\textwidth]{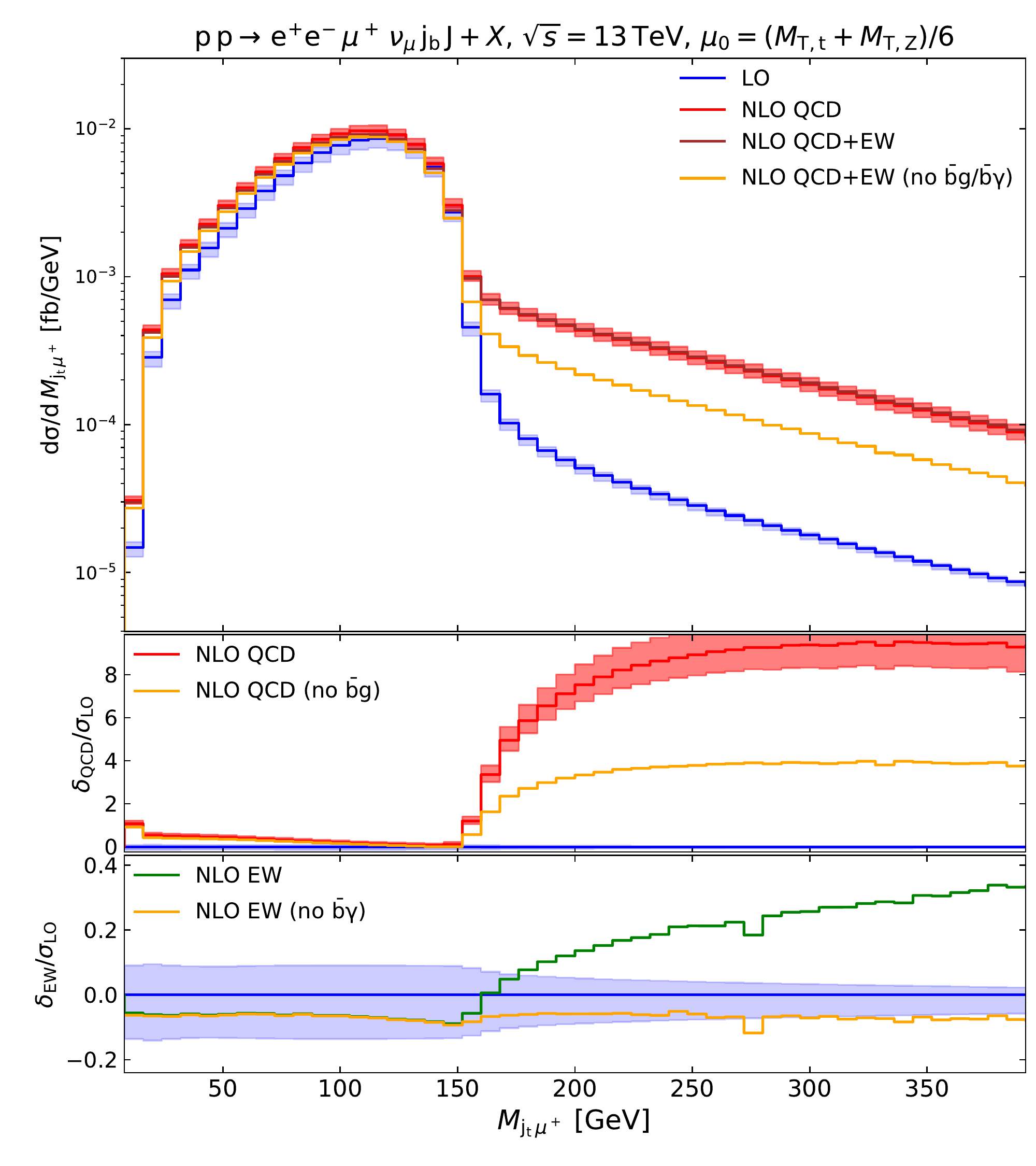}}\hspace{0.01\textwidth}%
\subfigure[\label{fig:cosine}]{\includegraphics[width=0.49\textwidth]{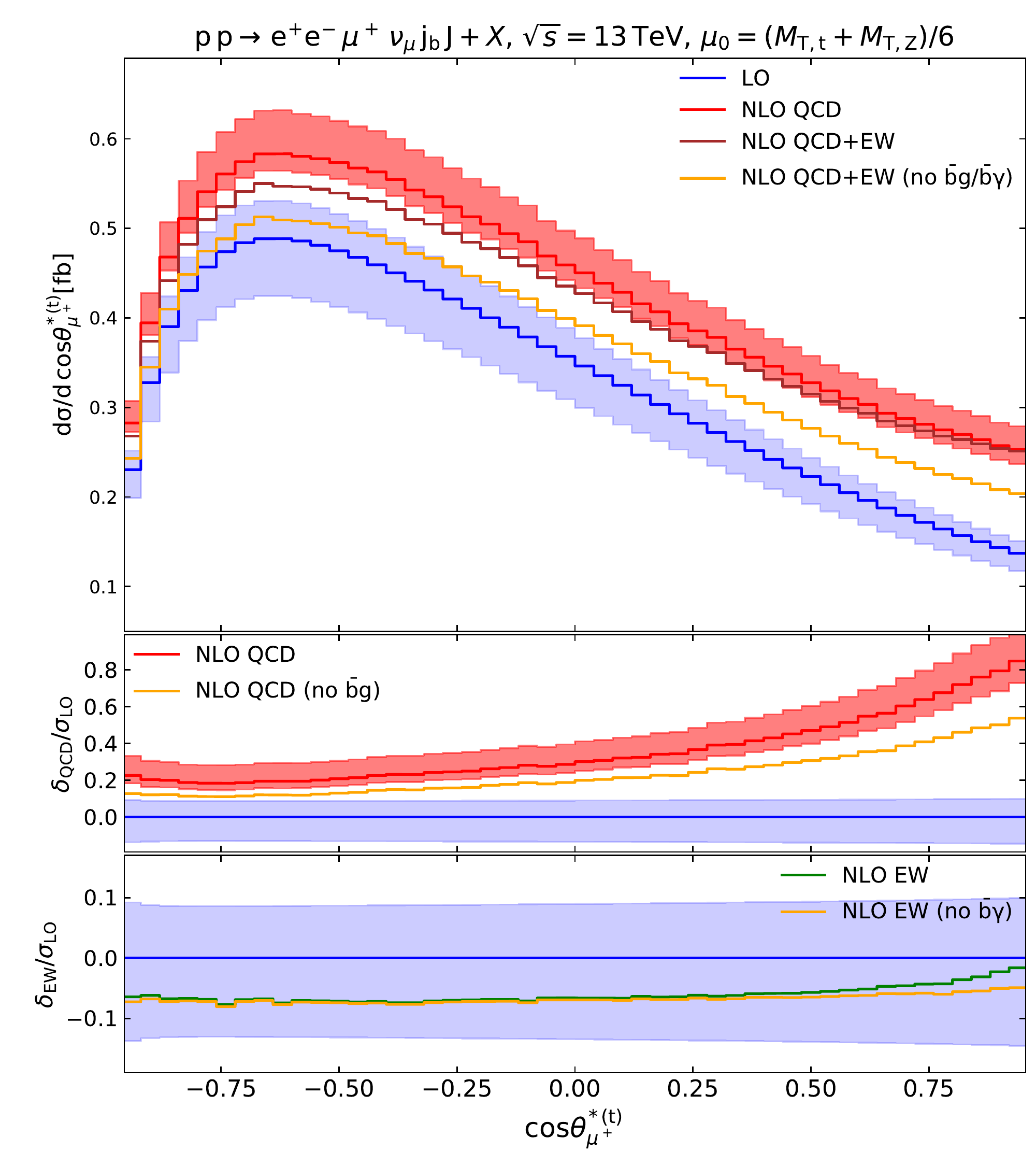}}
\caption{Invariant-mass distribution of the top-decay jet and the
  anti-muon, and distribution of the cosine of the angle of the anti-muon
   with respect to the direction of the
  boost from the laboratory frame to the top-quark rest frame in the latter frame.
The definition of the angle, given in Eq.~\eqref{eq:cosine-definition}, makes uses of the kinematic reconstruction routine given in \refse{sec:kinreco}.}
\label{fig:miscellaneous}
\end{figure}

Figure \ref{fig:mass-top-jet-anti-muon} shows the distribution of the
cross section in the invariant mass of the top-decay jet and the
anti-muon, which peaks at $M_{\mathrm{j}_\mathrm{t} \mu^+} \approx
\SI{120}{\giga\electronvolt}$ and $M_{\mathrm{j}_\mathrm{t} \mu^+}
\approx \SI{110}{\giga\electronvolt}$ in LO and NLO QCD, respectively,
displaying a sharp drop around $M_{\mathrm{j}_\mathrm{t} \mu^+} =
\sqrt{M_\mathrm{t}^2 - M_\mathrm{W}^2} \approx
\SI{153}{\giga\electronvolt}$, the threshold for having both an
on-shell top quark and W boson.
Above this threshold,
we find large corrections owing to
off-shell top-quark production and the $\bar\Pb\Pg/\bar\Pb\gamma$
anti-top production channels (both in NLO QCD and NLO EW), in which the W boson is not part of the top-quark decay.

Finally, in \reffi{fig:cosine}, we present the distribution in the cosine of the decay
angle of the anti-muon in the top-quark rest frame.
This angle is defined as the angular separation between the anti-muon
momentum in the top-quark rest frame, $\vec{p}^{\,*}_{\mu^+}$,
and the direction of the boost from the laboratory frame to the
top-quark rest frame, $\vec{n}^*_{\mathrm{t}}$, (top-quark momentum in
the laboratory frame)
\beq
\cos \theta_{\mu^+}^{\,* (\mathrm{t})}=
\frac{\vec{p}^{\,*}_{\mu^+}\cdot\vec{n}^*_{\mathrm{t}}}
{| \vec{p}^{\,*}_{\mu^+} | | \vec{n}^*_{\mathrm{t}} |} \,,
\label{eq:cosine-definition}
\eeq
where the top-quark momentum is reconstructed with the procedure detailed
in \refse{sec:kinreco}. This observable is highly sensitive to the
helicity of the top quark. In the absence of radiative corrections
and of cuts on the decay products, the (normalised) top-quark
decay rate for $\Pt\rightarrow \Pb\ell^+\nu_\ell$ reads
\beq\label{eq:topdecay}
\frac1{\sigma}\frac{\rd \sigma}{\rd \cos\theta^{*(\Pt)}_{\ell^+}} =
\frac12\left(
1+\kappa_{\ell^+}(f_+-f_-)\cos\theta^{*(\Pt)}_{\ell^+}
\right)\,,
\eeq
where $\kappa_{\ell^+}$ (=1 in the SM)  is the spin analyser of the charged lepton
and therefore sensitive to anomalous $\Pt\PW\Pb$
couplings \cite{Jezabek:1994zv,Aguilar-Saavedra:2006qvv,Aguilar-Saavedra:2014eqa},
while $f_+$, $f_-$ are the right- and left-helicity fractions dictated
by the dynamics of the top-quark production.
Due to the intrinsic frame dependence of helicity, different choices are possible for the quantisation axis
for the top-quark spin \cite{Mahlon:1999gz,Schwienhorst:2010je,Aguilar-Saavedra:2012bvs}.
In single-top-production studies usual choices are
the so-called helicity basis \cite{Jezabek:1994zv,Schwienhorst:2010je} and the
spectator basis \cite{Mahlon:1999gz,Schwienhorst:2010je,Aguilar-Saavedra:2014eqa}.
The angle defined in Eq.~\eqref{eq:cosine-definition} probes top-quark
helicities defined with respect to the direction of the boost from the
laboratory frame to the top-quark rest frame, which is a simple choice at the LHC and gives
similar results as the helicity basis \cite{Schwienhorst:2010je}.
Although used in experimental analyses \cite{CMS:2021ugv}, the spectator-basis
(helicities defined with the spectator-jet direction as reference axis) is a less natural choice
in the presence of an additional $\PZ$ boson that recoils against the system formed by the top quark
and the spectator jet.

Even though the shape expected from Eq.~\eqref{eq:topdecay} is distorted by fiducial cuts and
kinematic reconstruction of the top quark, resulting in the depletion of the anti-collinear
region, the LO distribution shown in the top panel of \reffi{fig:cosine} clearly
shows a strong left-handed polarisation of the top quark. The NLO QCD corrections increase
towards large $\cos\theta^{*(\Pt)}_{\mu^+}$, with large contributions from gluon-induced partonic channels,
suggesting spin configurations in the real corrections that favour a right-handed top quark.
The EW corrections are flat for $\cos\theta^{*(\Pt)}_{\mu^+}\lesssim0.5$
but are reduced for larger values of  $\cos\theta^{*(\Pt)}_{\mu^+}$ in
particular also owing to contributions of the $\bar\Pb\gamma$ channels.

\section{Conclusions}\label{concl}
We have presented a calculation of NLO EW and QCD corrections to
off-shell $\Pt\PZ\Pj$ production at the LHC. All  effects of
non-resonant top-quark or Z-boson contributions 
and full spin correlations are accounted for both at LO and at NLO.
A realistic fiducial phase-space region is considered for the
Monte Carlo predictions of integrated and differential cross sections.

At integrated level, the NLO QCD and EW corrections amount
to $+31\%$ and $-7\%$, respectively, of the LO fiducial cross section.
In more exclusive and off-shell regions, the NLO QCD effects become of order
$+100\%$ and larger and generate strong shape distortions of LO distributions.
The NLO EW corrections reach the order of $-20\%$ in the
tails of several transverse-momentum distributions.

The considered inclusive fiducial volume enables the
opening of partonic channels embedding a $\bar{\Pt}\PW\PZ$ resonance
structure at NLO which gives a sizeable enhancement to the LO
cross section ($+12\%$ at integrated level). Vetoes on additional
jets and kinematic constraints may be considered in experimental
analyses to suppress this background.

The top-quark-decay effects with full off-shell matrix elements, so
far missing in the literature, represent a necessary ingredient for
the complete fixed-order (NLO) modelling of $\Pt\PZ\Pj$ production in
the SM. Therefore the results shown in this work will be relevant
for upcoming LHC differential measurements of this process.

\section*{Acknowledgements}
The authors would like to thank Mathieu Pellen for useful discussions
in the early stages of this project, as well as Jean-Nicolas Lang and Sandro
Uccirati for maintaining \recola. This work is supported by the German
Federal Ministry for Education and Research (BMBF) under contract
no.~05H21WWCAA and by the German Research Foundation (DFG) under
reference number DE~623/6-2.

\bibliographystyle{JHEPmod}
\bibliography{tzj}

\end{document}

%% file: macros.tex
\def\refeq#1{\mbox{(\ref{#1})}}
\def\reffi#1{\mbox{Fig.~\ref{#1}}}
\def\reffis#1{\mbox{Figs.~\ref{#1}}}
\def\refta#1{\mbox{Table~\ref{#1}}}

\def\refse#1{\mbox{Sect.~\ref{#1}}}

\def\citere#1{\mbox{Ref.~\cite{#1}}}
\def\citeres#1{\mbox{Refs.~\cite{#1}}}

\newcommand{\newc}{\newcommand}
\newc{\beq}{\begin{equation}}
\newc{\eeq}{\end{equation}}
\newc{\bit}{\begin{itemize}}
\newc{\eit}{\end{itemize}}
\newc{\ben}{\begin{enumerate}}
\newc{\een}{\end{enumerate}}
\newc{\bce}{\begin{center}}
\newc{\ece}{\end{center}}
\newc{\bfi}{\begin{figure}}
\newc{\efi}{\end{figure}}

\newcommand{\rd}{\mathrm d}

\newcommand{\rT}{{\mathrm{T}}}

\newcommand{\ie}{i.e.\ }

\newcommand{\MeV}{\ensuremath{\,\text{MeV}}\xspace}
\newcommand{\GeV}{\ensuremath{\,\text{GeV}}\xspace}
\newcommand{\TeV}{\ensuremath{\,\text{TeV}}\xspace}

\newcommand{\PH}{\ensuremath{\text{H}}\xspace}
\newcommand{\Pj}{\ensuremath{\text{j}}\xspace}
\newcommand{\Pp}{\ensuremath{\text{p}}}
\newcommand{\Pe}{\ensuremath{\text{e}}\xspace}
\newcommand{\Pb}{\ensuremath{\text{b}}\xspace}

\newcommand{\Pt}{\ensuremath{\text{t}}\xspace}
\newcommand{\Pu}{\ensuremath{\text{u}}\xspace}
\newcommand{\Pd}{\ensuremath{\text{d}}\xspace}

\newcommand{\Pg}{\ensuremath{\text{g}}}

\newcommand{\PW}{\ensuremath{\text{W}}\xspace}
\newcommand{\PZ}{\ensuremath{\text{Z}}\xspace}

\newcommand{\Mt}{\ensuremath{m_\Pt}\xspace}
\newcommand{\MH}{\ensuremath{M_\PH}\xspace}

\newcommand{\MW}{\ensuremath{M_\PW}\xspace}

\newcommand{\MZ}{\ensuremath{M_\PZ}\xspace}

\newcommand{\Gt}{\ensuremath{\Gamma_\Pt}\xspace}
\newcommand{\GH}{\ensuremath{\Gamma_\PH}\xspace}

\newcommand{\MSbar}{\ensuremath{\overline{{\text{MS}}\xspace}}}

\newcommand{\recola}{{\sc Recola}\xspace}

\newcommand{\collier}{{\sc Collier}\xspace}

\newcolumntype{.}{D{.}{.}{-1}}
\newcolumntype{d}[1]{D{.}{.}{#1}}
\colorlet{tableoverheadcolor}{gray!37.5}
\colorlet{tableheadcolor}{gray!25}
\colorlet{tablerowcolor}{gray!12.5}

\def\draftdate{\relax}
\def\mda{\relax}
\def\mua{\relax}
\def\mla{\relax}
\def\draft{
\def\thtystars{******************************}
\def\sixtystars{\thtystars\thtystars}
\typeout{}
\typeout{\sixtystars**}
\typeout{* Draft mode!
         For final version remove \protect\draft\space in source file *}
\typeout{\sixtystars**}
\typeout{}
\def\draftdate{\today}
\def\mua{\marginpar[\boldmath\hfil$\uparrow$]%
                   {\boldmath$\uparrow$\hfil}\color{black}%
                    \typeout{marginpar: $\uparrow$}\ignorespaces}
\def\mda{\color{red}\marginpar[\boldmath\hfil$\downarrow$]%
                   {\boldmath$\downarrow$\hfil}%
                    \typeout{marginpar: $\downarrow$}\ignorespaces}
\def\mla{\marginpar[\boldmath\hfil$\rightarrow$]%
                   {\boldmath$\leftarrow $\hfil}%
                    \typeout{marginpar: $\leftrightarrow$}\ignorespaces}
\def\Mua{\marginpar[\boldmath\hfil$\Uparrow$]%
                   {\boldmath$\Uparrow$\hfil}\color{black}%
                    \typeout{marginpar: $\uparrow$}\ignorespaces}
\def\Mda{\color{red}\marginpar[\boldmath\hfil$\Downarrow$]%
                   {\boldmath$\Downarrow$\hfil}%
                    \typeout{marginpar: $\downarrow$}\ignorespaces}
\def\Mla{\marginpar[\boldmath\hfil\textcolor{red}{$\Rightarrow$}]%
                   {\boldmath\textcolor{red}{$\Leftarrow $}\hfil}%
                    \typeout{marginpar: $\leftrightarrow$}\ignorespaces}
\overfullrule 5pt
\oddsidemargin 15mm
\marginparwidth 29mm
}

\newcommand{\mc}{\mathcal}

\let\Mwo\MWOS
\let\Gwo\GWOS
\let\Mzo\MZOS
\let\Gzo\GZOS

\let\as\alphas
\newcommand{\pt}[1]{p_{\rT,{#1}}}

\newcommand{\tj}{{\rm j}_{\rm t}}
\newcommand{\sj}{{\rm j}_{\rm s}}